\documentclass[english,prd,a4paper,onecolumn,nofootinbib]{revtex4}
\usepackage{graphicx, bm, color, babel}
\usepackage{hyperref}
\usepackage{amsmath}
\usepackage{amssymb}
\usepackage{amsfonts}
\usepackage{dsfont}
\usepackage{color}

\def\be{\begin{equation}}
\def\ee{\end{equation}}
\def\bea{\begin{eqnarray}}
\def\eea{\end{eqnarray}}

\newcommand{\HH}{\mathcal{H}}

\newcommand\lsim{\mathrel{\rlap{\lower4pt\hbox{\hskip1pt$\sim$}}
        \raise1pt\hbox{$<$}}}

\newcommand{\bn}{\mathbf{n}}
\newcommand{\bk}{\mathbf{k}}

\newcommand{\bx}{\mathbf{ x}}
\newcommand{\by}{\mathbf{y}}

\newcommand\spart{\;\raise1.0pt\hbox{$/$}\hskip-6pt\partial}
\newcommand\spartb{\;\overline{\raise1.0pt\hbox{$/$}\hskip-6pt
\partial}}

\newcommand{\B}{{\rm B}}
\newcommand{\F}{{\rm F}}
\newcommand{\w}[2]{w_{\bx_{#1}\bx_{#2}L_{#1} L_{#2}}}
\newcommand{\dn}{d\bar{n}}
\newcommand{\R}{\mathbf{V}\cdot\mathbf{n}}
\newcommand{\bv}{\mathbf{V}}
\newcommand{\unit}{1\!\!1}

\newcommand{\norm}{c_N}

\usepackage{hyperref}

\begin{document}

\title{\vspace{-0.5cm}{\normalsize \normalfont \flushright CERN-PH-TH-2015-220\\}
\vspace{0.6cm}
Optimising the measurement of relativistic distortions in large-scale structure}
\author{Camille Bonvin$^{1, 2}$}
\author{Lam Hui$^3$}
\author{Enrique Gaztanaga$^4$}
\affiliation{$^1$CERN, Theory Division, 1211 Geneva, Switzerland.\\
$^2$D\'epartement de Physique Th\'eorique and Center for Astroparticle Physics (CAP), University of Geneva,
24 quai Ernest Ansermet, CH-1211 Geneva, Switzerland.\\
$^3$Institute for Strings, Cosmology and Astroparticle Physics
and Department of Physics, 
Columbia University, New York, NY 10027, U.S.A.\\
$^4$Institut de Ci\`encies de l'Espai (IEEC-CSIC),
F. Ci\`encies, C5 2-par,
 Bellaterra,  Barcelona 08193, Spain.\\
 {\rm E-mail: camille.bonvin@unige.ch, lhui@astro.columbia.edu, gazta@ice.cat}}  \vspace*{0.2cm}

\date{\today}

\begin{abstract} 

It has been shown recently that relativistic distortions generate a dipolar modulation in the two-point correlation function of galaxies. To measure this relativistic dipole it is necessary to cross-correlate different populations of galaxies with for example different luminosities or colours. In this paper, we construct an optimal estimator to measure the dipole with multiple populations. We show that this estimator increases the signal-to-noise of the dipole by up to 35 percent. Using 6 populations of galaxies, in a survey with halos and number densities similar to those of the millennium simulation, we forecast a cumulative signal-to-noise of 4.4. For the main galaxy sample of SDSS at low redshift $z\leq 0.2$ our optimal estimator predicts a cumulative signal-to-noise of 2.4. Finally we forecast a cumulative signal-to-noise of 7.4 in the upcoming DESI survey. These forecasts indicate that with the appropriate choice of estimator the relativistic dipole should be detectable in current and future surveys.

\end{abstract}

\maketitle

\section{Introduction}

The two-point correlation function of galaxies, and its Fourier transform the power spectrum, have been measured with increasing precision over the last decades. These observables contain valuable information about the global properties of our universe: its initial conditions, the law of gravity governing its evolution and the amount and nature of dark energy. To a good approximation, the observed two-point correlation function of galaxies follows the two-point correlation function of the underlying dark matter. The galaxy correlation function shares therefore the same statistical properties as the dark matter correlation function. In particular this implies that the galaxy correlation function is isotropic and that its shape depends only on the galaxies' separation. The relation between the galaxy and the dark matter correlation function is simply given via the square of the bias, which in the linear regime is usually assumed to be scale independent. In this context, measuring the galaxy correlation function allows to directly characterise the distribution of dark matter in our universe.

Since the 80's we know however that this description of the two-point function of galaxies is too simplistic since it does not account for the fact that our observations are made in redshift-space~\cite{kaiser,lilje,hamilton}. In redshift-space, the peculiar velocities of galaxies distort the two-point correlation function. As a consequence the correlation function is not isotropic anymore: it depends on the orientation of the pair of galaxies with respect to the observer's line-of-sight. One can show that in the distant-observer approximation, redshift distortions generate a quadrupole and an hexadecapole.
Measurements of these multipoles have been performed in various galaxy surveys (see e.g.~\cite{hawkins, zehavi, guzzo, cabre, song, bossCMASS, chuang}). These measurements provide additional information on our universe since they are sensitive to the galaxies' peculiar velocities. In particular, combined measurements of the monopole and of the quadrupole allow to measure separately the bias and the growth rate of fluctuations~\footnote{More precisely, one can measure separately $b\sigma_8$ and $f\sigma_8$.}. The fact that we observe in redshift-space is therefore not a complication, but rather an interesting source of information.

In the past few years, it has been shown that redshift-space distortions are just one of the many distortions that affect the observed distribution of galaxies. The fractional over-density of galaxies $\Delta$ is distorted by gravitational lensing~\cite{gunn, turner, webster, fugmann, narayan, schneider, broadhurst, moessner, loverde}, Doppler effects, gravitational redshift, Sachs-Wolfe effects, Shapiro time-delay and integrated Sachs-Wolfe~\cite{yoo, galaxies, challinor, schmidt}. These terms distort the coordinate system in which our observations are performed (i.e. redshift and incoming photon's direction) and they generate consequently additional fluctuations to $\Delta$. These effects (apart from lensing and Doppler effects) have been called {\it relativistic distortions}, since they are suppressed by powers of $\HH/k$ with respect to the standard contributions --namely density and redshift-space distortions-- and they are therefore expected to become relevant at large scales, near the horizon. A natural question arises then: can we use similar techniques as those developed for redshift-space distortions to {\it isolate} the relativistic distortions from the standard terms? 

In a recent paper~\cite{assym} (see also~\cite{roy,croft,review}), we showed that some of the relativistic distortions have a remarkable property: they break the symmetry of the correlation function under the exchange of the two galaxies in the pair (this property has been identified previously in Fourier space, where the relativistic distortions generate an imaginary part to the power spectrum~\cite{mcdonald,yoo_im}). Obviously to observe such a breaking of symmetry we need more that one population of galaxies. In~\cite{assym}, we showed that the cross-correlation function between a bright and a faint population of galaxies contains in addition to the standard monopole, quadrupole and hexadecapole, a {\it dipole} and an {\it octupole} directly generated by the relativistic distortions. This suggests that we can isolate the contributions from relativistic distortions by fitting for a dipole and an octupole in the two-point function. Since these new multipoles are orthogonal to the monopole, quadrupole and hexadecapole, this method would allow us to get rid of the dominant standard terms and to target specifically the new relativistic terms. In essence this is very similar to the method used in~\cite{Wojtak,Lahav} to measure gravitational redshift in clusters and to separate it from the dominant Doppler redshift. 

In this paper, we calculate the detectability of the relativistic distortions in large-scale structure using this method. We construct an optimal estimator to isolate the dipole. We then calculate the signal-to-noise for the dipole in a multi-population case and we show that our optimal estimator allows us to improve the measurement of the relativistic distortions by up to 35 percent. In a survey with halos and number densities similar to those of the millennium simulation we expect a cumulative signal-to-noise of 4.4. For the main galaxy sample of SDSS (DR5) at low redshift we can reach a cumulative signal-to-noise of 2.4. Finally we forecast a cumulative signal-to-noise of 7.4 in the upcoming Dark Energy Spectroscopic Instrument (DESI) survey. This demonstrates the feasibility of our method to current and future surveys. The advantage of this method is that it does not require to measure the correlation function at extremely large scales, of the size of the horizon. By fitting for a dipole we can indeed isolate the relativistic distortions from the standard terms at scales accessible by current surveys. 

The remainder of the paper is organised as follow: in section~\ref{sec:multipop} we construct a general estimator combining different populations of galaxies, which isolates the anti-symmetric part of the correlation function. In section~\ref{sec:variance} we derive the variance of this estimator. In section~\ref{sec:opt} we find the kernel which minimises the variance. In section~\ref{sec:signaltonoise} we calculate the signal-to-noise of the dipole and in section~\ref{sec:measurement} we forecast our method to the millennium simulation, the main galaxy sample of SDSS and the DESI survey.

\section{The two-point correlation function for multiple populations of galaxies}
\label{sec:multipop}

To measure anti-symmetric terms in the two-point correlation function we need more than one population of galaxies. If all galaxies are the same, we have indeed by construction that $\langle \Delta(\bx, z)\Delta(\bx', z')\rangle$
is symmetric under the exchange of the two galaxies in the pair. If however we split the galaxies into multiple populations with different characteristics, e.g. different luminosities, then the cross-correlation between two populations with respective luminosities $L$ and $L'$ can have an anti-symmetric part
\be
\langle \Delta_L(\bx, z)\Delta_{L'}(\bx', z')\rangle\neq\langle \Delta_L(\bx', z')\Delta_{L'}(\bx, z)\rangle\, .
\ee 

The goal of this paper is to construct an estimator which isolates the anti-symmetric part of the correlation function. 
We start by splitting the galaxies depending on their luminosity. For each pixel $i$ in the sky, we then count how many galaxies we have at each luminosity (or in each bin of luminosity). Let us denote this number by $n_{L_i}(\bx_i)$, where $L_i$ is the luminosity of the population under consideration and $\bx_i$ is the position of the pixel~\footnote{Here and in the following we work in the distant-observer approximation and we neglect the evolution of our observables with redshift (see text after eq.~\eqref{Deltarel_Euler}). All variables are therefore evaluated at the mean redshift of the survey $\bar z$. For example $n_{L_i}(\bx_i, z_i)\simeq n_{L_i}(\bx_i, \bar z)$. For simplicity we drop the dependence on $\bar z$ in the notation, when it is not needed.}. The overdensity of galaxies in pixel $i$ with luminosity $L_i$ is then
\be
\delta n_{L_i}(\bx_i)=n_{L_i}(\bx_i)-\dn_{L_i}\,,
\ee
where $\dn_{L_i}$ denotes the mean number of galaxies per pixel, with luminosity $L_i$. Note that $\dn_{L_i}$ depends on the size of the pixels. The most general estimator we can construct, combining all populations of galaxies, is then~\footnote{Note that in the estimator we do not divide $\delta n_{L_i}(\bx_i)$ by $\dn_{L_i}$, as is usually done for auto-correlations. The reason is that $\dn_{L_i}$ depends on luminosity and dividing by this factor is similar to applying a weighting to the different populations. Such a factor should therefore be included in the general kernel $\w{i}{j}$.}
\be
\label{estimator}
\hat\xi=\sum_{ij}\sum_{L_i L_j} w_{\bx_i \bx_j L_i L_j}\delta n_{L_i}(\bx_i)\delta n_{ L_j}(\bx_j)\, ,
\ee
where the kernel $w_{\bx_i \bx_j L_i L_j}$ depends on the position of the pixels $i$ and $j$ and on the luminosities $L_i$ and $L_j$. This kernel must be symmetric under the exchange of $i$ and $j$, which just represents a relabelling of the pixels
\be
\label{wsym}
w_{\bx_i \bx_j L_i L_j}= w_{\bx_j \bx_i L_j L_i}\, .
\ee

We want to construct a kernel which isolates the anti-symmetric part of the correlation function. The general expression for the overdensity of galaxies reads
\be
\label{Deltatot}
\delta n_{L_i}(\bx_i)=\dn_{L_i}\cdot\Delta_{L_i}(\bx_i)=\dn_{L_i}\left[b_{L_i}\delta_i-\frac{1}{\HH}\partial_r(\bv\cdot\bn)_i +
(5s_{L_i}-2)\int_0^{r_i}\!dr\,\frac{r_i-r}{2rr_i}\Delta_\Omega(\Phi+\Psi)+\Delta_{L_i}^{\rm rel}(\bx_i)\right]\, ,
\ee
where $\Phi$ and $\Psi$ are the two metric potentials~\footnote{We use here the following convention for the metric $ds^2=a^2\big[-(1+2\Psi)d\eta^2+(1-2\Phi)\delta_{ij}dx^i dx^j \big]$, where $a$ is the scale factor and $\eta$ denotes conformal time.}, $\delta$ is the density contrast, $\bv$ is the peculiar velocity, $\bn$ is the observed direction, $\HH$ is the conformal Hubble parameter, $r$ is the conformal distance to the source and $\Delta_\Omega$ denotes the angular Laplacian. The indices $i$ represents a value evaluated in pixel $i$, and $b_{L_i}$ and $s_{L_i}$ denote respectively the bias and the slope of the luminosity function of the galaxy population with luminosity $L_i$.  
The first term in eq.~\eqref{Deltatot} is the density contribution and the second term is the contribution from redshift-space distortions. We call the sum of these two contributions the {\it standard} terms
\be
\label{delta_stand}
\delta n^{\rm stand}_{L_i}(\bx_i)=\dn_{L_i}\left[b_{L_i}\delta_i-\frac{1}{\HH}\partial_r (\bv\cdot\bn)_i\right]\, .
\ee
The third term in eq.~\eqref{Deltatot} denotes the so-called lensing magnification bias which depends on the slope of the luminosity function $s_{L_i}$ and the last term encodes all the relativistic distortions.  The relativistic distortions contain contributions with one gradient of the gravitational potentials and contributions directly proportional to the potentials. As shown in~\cite{assym}, the terms with one gradient of the potentials are those that generate an anti-symmetry in the correlation function~\footnote{Note that the lensing magnification bias also generates an anti-symmetry in the two-point function. However, as shown in~\cite{assym} this contribution is always significantly smaller than the terms in eq.~\eqref{Deltarel} and it can safely be neglected.}. They read
\be
\label{Deltarel}
\delta n_{L_i}^{\rm rel}(\bx_i)=\dn_{L_i}\left\{\frac{1}{\HH}\partial_r\Psi_i + \frac{1}{\HH}(\dot{\bv}\cdot\bn)_i
+\left[1-\frac{\dot\HH}{\HH^2}-\frac{2}{r\HH}+5s_{L_i}\left(1-\frac{1}{r\HH} \right)\right](\bv\cdot\bn)_i\right\}\, ,
\ee 
where the first term is the contribution from gravitational redshift and the other terms are Doppler contributions. Here a dot denotes derivative with respect to conformal time $\eta$. In theories of gravity where Euler equation is valid, we can rewrite the gradient of the potential as a function of velocity and we obtain
\be
\label{Deltarel_Euler}
\delta n_{L_i}^{\rm rel}(\bx_i)=
-\dn_{L_i}\left(\frac{\dot\HH}{\HH^2}+\frac{2}{r\HH} \right)(\bv\cdot\bn)_i\, .
\ee 
Here and in the following we neglect for simplicity the contribution proportional to the slope of the luminosity function in eq.~\eqref{Deltarel}, i.e. we set $s_{L_i}=0$. We also neglect the anti-symmetric contributions generated by evolution effects and wide-angle effects, assuming that all the anti-symmetric signal is due to the relativistic distortions in eq.~\eqref{Deltarel_Euler}. The effect of evolution has been calculated in~\cite{assym} and shown to be small: less than 5\% of the relativistic signal at redshift $z=0.25$ and less than 9\% at $z=0.5$ (see blue dashed line of figure 11). The wide-angle effects have been calculated in a companion paper~\cite{kernel_obs}, where we show that with the appropriate choice of kernel they are of the same order of magnitude as the relativistic effects (albeit slightly smaller) and with an opposite sign, see figure 3 of~\cite{kernel_obs} left panel. In the following we disregard nevertheless the wide-angle effects. We are indeed primarily interested to find the kernel that will optimise the measurement of the relativistic dipole. As discussed in~\cite{assym}, once the dipole has been measured we can in principle easily separate the wide-angle contribution from the relativistic contribution by using measurements of the quadrupole. Another strategy would be to find the kernel that optimise the measurement of the relativistic dipole while minimising the contribution from the wide-angle dipole. We defer this more involved calculation to a future paper.

\begin{figure}[t]
\centering
\includegraphics[width=0.19\textwidth]{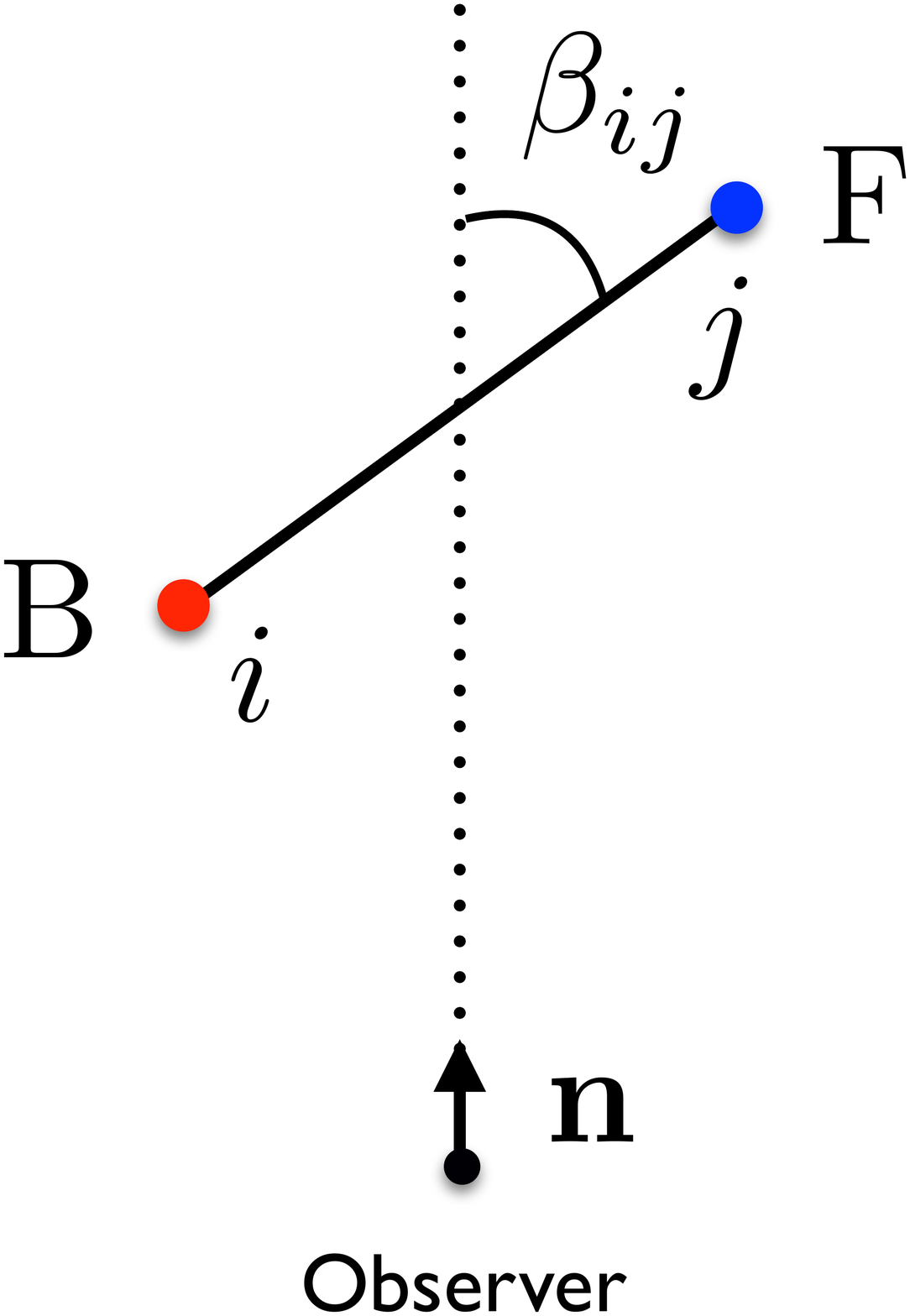}
\caption{\label{fig:coordinate} Coordinate system in which the dipole is observed.}
\end{figure}

As shown in~\cite{assym}, in the case of two populations of galaxies --a bright and a faint population-- the contribution~\eqref{Deltarel_Euler} generates a dipole in the correlation function 
\begin{align}
\label{dipole}
&\langle \delta n_\B^{\rm stand}(\bx_i)\delta n_\F^{\rm rel}(\bx_j)\rangle+\langle \delta n_\F^{\rm rel}(\bx_i)\delta n_\B^{\rm stand}(\bx_j)\rangle=\nonumber\\
&d\bar n_\B d\bar n_\F(b_\B-b_\F)\left(\frac{\dot\HH}{\HH^2}+\frac{2}{r\HH} \right)\frac{\HH}{\HH_0}\frac{f}{2\pi^2}
\int dk k \HH_0 P(k, \bar z)j_1(kd_{ij})\cdot\cos\beta_{ij} \, ,
\end{align}
where $P(k, \bar z)$ is the density power spectrum at the mean redshift of the survey:
\be
\langle \delta(\bk, \bar z)\delta(\bk', \bar z) \rangle=(2\pi)^3 P(k, \bar z)\delta_D(\bk+\bk')\, ,
\ee
$f$ is the growth rate, $d_{ij}$ is the pair separation and $\beta_{ij}$ denotes the orientation of the pair with respect to the line-of-sight~\footnote{Note that since we work here in the distant-observer approximation, the line-of-sight to the median, to the bright and to the faint galaxy are all parallel.}, as depicted on figure~\ref{fig:coordinate}.

Eq.~\eqref{dipole} is anti-symmetric under the exchange of the bright and faint population: $\dn_\B\dn_\F(b_\B-b_\F)=-\dn_\F\dn_\B(b_\F-b_\B)$. It is also anti-symmetric under the exchange of the relative position of the pixels, i.e. when $\beta_{ij}$ goes into $\beta_{ij}+\pi$. To isolate this term, we need therefore a kernel $w$ which is anti-symmetric under the exchange of $L_i$ and $L_j$
\be
\label{antiL}
w_{\bx_i \bx_j L_i L_j}= -w_{\bx_i \bx_j L_j L_i}\, ,
\ee
as well as anti-symmetric under the exchange of $\bx_i$ and $\bx_j$
\be
\label{antix}
w_{\bx_i \bx_j L_i L_j}= -w_{\bx_j \bx_i L_i L_j}\, .
\ee

These properties of the kernel insure us that the standard density and redshift-space distortions do not contribute to the mean of the estimator. Inserting~\eqref{delta_stand} into~\eqref{estimator} we have indeed
\begin{align}
\big\langle \hat\xi^{\rm stand} \big\rangle=\sum_{ij}\sum_{L_i L_j} w_{\bx_i \bx_j L_i L_j}\dn_{L_i}\dn_{L_j}&\Big[b_{L_i}b_{L_j}\langle \delta_i \delta_j\rangle
-\frac{1}{\HH}b_{L_i}\langle\delta_i\,\partial_r (\bv\cdot\bn)_j\rangle -\frac{1}{\HH}b_{L_j}\langle\delta_j\,\partial_r (\bv\cdot\bn)_i\rangle\nonumber\\
&+\frac{1}{\HH^2}\langle\partial_r (\bv\cdot\bn)_i\partial_r (\bv\cdot\bn)_j\rangle\Big]=0\, ,\label{mean_st}
\end{align}
since $w_{\bx_i \bx_j L_i L_j}$ is anti-symmetric under the exchange of $L_i$ and $L_j$ whereas the bracket in~\eqref{mean_st} is symmetric (to show this we use that $\langle\delta_j\,\partial_r (\bv\cdot\bn)_i\rangle=\langle\delta_i\,\partial_r (\bv\cdot\bn)_j\rangle$).
This argument applies also to all relativistic distortions in $\Delta_{L_i}^{\rm rel}$ that have no gradients of $\Phi$ and $\Psi$. On the other hand, the terms in eq.~\eqref{Deltarel_Euler}, which have only one gradient of the potential, survive in the mean
\bea
\label{xirel}
\big\langle \hat\xi \big\rangle=\big\langle \hat\xi^{\rm rel} \big\rangle&=&-\left(\frac{\dot\HH}{\HH^2}+\frac{2}{r\HH} \right)\sum_{ij}\sum_{L_i L_j} w_{\bx_i \bx_j L_i L_j}\dn_{L_i}\dn_{L_j}\Big[b_{L_i}\big\langle \delta_i (\R)_j\big\rangle
+b_{L_j}\big\langle \delta_j (\R)_i\big\rangle\Big]\nonumber\\
&=&-\left(\frac{\dot\HH}{\HH^2}+\frac{2}{r\HH} \right)\sum_{ij}\sum_{L_i L_j} w_{\bx_i \bx_j L_i L_j}\dn_{L_i}\dn_{L_j}\big(b_{L_i}-b_{L_j}\big)\big\langle \delta_i (\R)_j\big\rangle\,,
\eea
where we have used that 
\bea
\langle \delta_i (\R)_j\big\rangle&=& -\frac{\HH}{\HH_0}\frac{f}{2\pi^2}
\int dk k \HH_0 P(k, \bar z)j_1(kd_{ij})\cdot \cos(\beta_{ij})\\
&=&\frac{\HH}{\HH_0}\frac{f}{2\pi^2}
\int dk k \HH_0 P(k, \bar z)j_1(kd_{ji})\cos(\beta_{ji})=-\big\langle \delta_j (\R)_i\big\rangle\, . \nonumber
\eea
Since $\dn_{L_i}\dn_{L_j}\big(b_{L_i}-b_{L_j}\big)$ is clearly anti-symmetric under the exchange of $L_i$ and $L_j$, eq.~\eqref{xirel} does not vanish under the summation over $L_i$ and $L_j$. The generic kernel defined by eqs.~\eqref{antiL} and~\eqref{antix} therefore allows us to isolate the relativistic terms from the dominant density and redshift-space distortions. 

A simple example for the kernel $w$ is 
\be
\label{kernel1}
\w{i}{j}=\frac{3}{8\pi}\Big[\theta(L_i-L_j) -\theta(L_j-L_i) \Big] \cos\beta_{ij}\delta_K(d_{ij}-d)\, ,
\ee
where $\theta$ is the Heaviside function.  
In the case of two populations of galaxies (bright and faint), inserting kernel~\eqref{kernel1} into eq.~\eqref{xirel} we obtain
\be
\label{xirelBF}
\langle \hat\xi \rangle
=\norm\dn_{\B}\dn_{\F}\big(b_{\B}-b_{\F}\big)\left(\frac{\dot\HH}{\HH^2}+\frac{2}{r\HH} \right)\frac{\HH}{\HH_0}\frac{f}{2\pi^2}
\int dk k \HH_0 P(k, \bar z)j_1(kd)\,,
\ee
where $\norm$ is a normalisation factor.

We now want to find the kernel which maximises the signal-to-noise of the dipole in a generic survey with multiple populations of galaxies.

\section{Variance}
\label{sec:variance}

We start by calculating the variance of our estimator~\eqref{estimator}:
\begin{align}
{\rm var}(\hat\xi)=\langle \hat\xi^2\rangle-\langle \hat\xi\rangle^2
=&\sum_{ijL_iL_j}\sum_{ab L_aL_b}\w{i}{j}\w{a}{b}\nonumber\\
&\times\Big(\big\langle\delta n_{L_i}(\bx_i)\delta n_{L_j}(\bx_j)\delta n_{L_a}(\bx_a)\delta n_{L_b}(\bx_b)\big\rangle
-\big\langle\delta n_{L_i}(\bx_i)\delta n_{L_j}(\bx_j)\big\rangle\big\langle\delta n_{L_a}(\bx_a)\delta n_{L_b}(\bx_b)\big\rangle\Big)\nonumber\\
=&2\sum_{ijL_iL_j}\sum_{ab L_aL_b}\w{i}{j}\w{a}{b}\label{var}
\big\langle\delta n_{L_i}(\bx_i)\delta n_{L_a}(\bx_a)\big\rangle\big\langle \delta n_{L_j}(\bx_j)\delta n_{L_b}(\bx_b)\big\rangle\, ,
\end{align}
where we have used Gauss theorem to expand the four-point correlation function in products of two-point functions and we have used the symmetry property of the kernel eq.~\eqref{wsym}.

\subsection{Poisson noise}

The first contribution to eq.~\eqref{var} comes from Poisson noise. We have
\bea
\big\langle\delta n_{L_i}(\bx_i)\delta n_{L_a}(\bx_a)\big\rangle &=&\big\langle \big(n_{L_i}(\bx_i)-d\bar n_{L_i}\big)\big(n_{L_a}(\bx_a)-d\bar n_{L_a}\big)\big\rangle\nonumber\\
&=&\big\langle n_{L_i}(\bx_i)n_{L_a}(\bx_a)\big\rangle-d\bar n_{L_i}d\bar n_{L_a}\, .
\eea
We have different cases:
\begin{itemize}
\item $i\neq a$: $\big\langle n_{L_i}(\bx_i)n_{L_a}(\bx_a)\big\rangle=\big\langle n_{L_i}(\bx_i)\big\rangle\big\langle n_{L_a}(\bx_a)\big\rangle=d\bar n_{L_i}d\bar n_{L_a}$, so that $\big\langle\delta n_{L_i}(\bx_i)\delta n_{L_a}(\bx_a)\big\rangle=0$.
\item $i=a$ and $L_i=L_a$: $\big\langle n_{L_i}^2(\bx_i)\big\rangle=d\bar n_{L_i}+d\bar n^2_{L_i}$, so that 
$\big\langle \delta n^2_{L_i}(\bx_i)\big\rangle=d\bar n_{L_i}$.
\item $i=a$ and $L_i\neq L_a=L'_i$, i.e. in the same pixel $i$ we look at different populations. The Poisson fluctuations of these different populations are uncorrelated so that
\be
\big\langle n_{L_i}(\bx_i)n_{L'_i}(\bx_i)\big\rangle=\big\langle n_{L_i}(\bx_i)\big\rangle\big\langle n_{L'_i}(\bx_i)\big\rangle\rangle
=d\bar n_{L_i}d\bar n_{L'_i}\hspace{1cm}\mbox{and}\hspace{1cm}\big\langle \delta n_{L_i}(\bx_i) \delta n_{L'_i}(\bx_i)\big\rangle=0\, .
\ee
\end{itemize}
The Poisson noise can therefore generally be written as
\be
\label{nnP}
\big\langle\delta n_{L_i}(\bx_i)\delta n_{L_a}(\bx_a)\big\rangle=d\bar n_{L_i} \delta_{ia}\delta_{L_iL_a}\, .
\ee
Inserting this in the variance we obtain 
\be
\label{varP}
{\rm var_P}(\hat\xi)=2\sum_{ijL_iL_j}(\w{i}{j})^2d\bar n_{L_i}d\bar n_{L_j}\, . 
\ee
Even if the kernel is anti-symmetric in $L_i\leftrightarrow L_j$, the Poisson term does not vanish because according to eq.~\eqref{nnP} it is non-zero only when $i=a$ {\it and} $L_i=L_a$. As a consequence, only the square of the kernel (which is symmetric in $L_i\leftrightarrow L_j$) enters into eq.~\eqref{varP}. Note that to derive~\eqref{varP} we have assumed that galaxies follow Poisson statistics. This is a simplifying assumption. Simulations have indeed shown that exclusion and non-linear clustering effects generate non-diagonal shot-noise contributions through correlations between galaxies of different luminosities and correlations between close pixels~\cite{baldauf}.

\subsection{Cosmic variance}

The second contribution to eq.~\eqref{var} comes from the cosmic variance of the density and redshift distortions contributions~\eqref{delta_stand} (which dominate over the cosmic variance of the relativistic terms)
\begin{align}
\label{var_cosmic}
{\rm var_C}(\hat\xi)=&2\sum_{ijab}\sum_{L_iL_jL_a L_b}d\bar n_{L_i}d\bar n_{L_j}d\bar n_{L_a}d\bar n_{L_b}\w{i}{j}\w{a}{b}\\
&\times\Big[ b_{L_i}b_{L_a}\langle\delta_i\delta_a\rangle-\frac{1}{\HH}b_{L_i}\langle\delta_i\,\partial_r(\bv\cdot\bn)_a\rangle
-\frac{1}{\HH}b_{L_a}\langle\delta_a\,\partial_r(\bv\cdot\bn)_i\rangle+\frac{1}{\HH^2}\langle\partial_r(\bv\cdot\bn)_i\partial_r(\bv\cdot\bn)_a \rangle\Big]\nonumber\\
&\times\Big[ b_{L_j}b_{L_b}\langle\delta_j\delta_b\rangle-\frac{1}{\HH}b_{L_j}\langle\delta_j\,\partial_r(\bv\cdot\bn)_b\rangle
-\frac{1}{\HH}b_{L_b}\langle\delta_b\,\partial_r(\bv\cdot\bn)_j\rangle+\frac{1}{\HH^2}\langle\partial_r(\bv\cdot\bn)_j\partial_r(\bv\cdot\bn)_b \rangle\Big]\, .\nonumber
\end{align}
Many of the products in the brackets vanish since they are symmetric under the exchange $L_i\leftrightarrow L_j$ or $L_a\leftrightarrow L_b$, whereas the kernels $\w{i}{j}$ and $\w{a}{b}$ are anti-symmetric. The remaining terms read
\begin{align}
\label{varCgen}
{\rm var_C}(\hat\xi)=&\sum_{ijab}\sum_{L_iL_jL_a L_b}d\bar n_{L_i}d\bar n_{L_j}d\bar n_{L_a}d\bar n_{L_b}\w{i}{j}\w{a}{b}
(b_{L_i}-b_{L_j})(b_{L_a}-b_{L_b})\\
&\times  \frac{1}{\HH^2}\Big[\langle\delta_i\delta_a \rangle \langle\partial_r(\mathbf{V}\cdot\mathbf{n})_j\partial_r(\mathbf{V}\cdot\mathbf{n})_b \rangle
- \langle\delta_i\partial_r(\mathbf{V}\cdot\mathbf{n})_a \rangle\langle\delta_j\partial_r(\mathbf{V}\cdot\mathbf{n})_b \rangle \Big]\, .\nonumber
\end{align}
As shown in appendix~\ref{a:variance}, going to the continuous limit and using a change of variables we can show that the second line of eq.~\eqref{varCgen} is proportional to
\begin{align}
\frac{1}{45}+\frac{2}{63}P_2(\bn\cdot\hat{\bk})+\frac{2}{35}P_4(\bn\cdot\hat{\bk})-\frac{1}{9}P_2^2(\bn\cdot\hat{\bk}) =0\, ,
\end{align}
where $P_\ell$ denotes the Legendre polynomial of degree $\ell$.
This shows that the properties of the kernel allow us to get rid of the standard dominant terms (density and redshift-space distortions) not only in the signal but also in the variance. This cancellation greatly enhances the detectability of the relativistic dipole. Note that the cancellation of the cosmic variance for multiple populations of galaxies has already been discussed in detail in the case of the power spectrum~\cite{seljak_tracers,mcdonald_tracers}.

\subsection{Mixed term}
\label{sec:mixed}

The variance~\eqref{var} also contains mixed contributions, where the cosmic variance contributes to one of the correlation and the Poisson noise to the other. These terms read
\begin{align}
\label{varCP}
{\rm var_{CP}}(\hat\xi)=&4\sum_{ija}\sum_{L_iL_jL_a}\w{i}{j}\w{a}{j}d\bar n_{L_i}d\bar n_{L_j}d\bar n_{L_a}\\
&\times\Big[ b_{L_i}b_{L_a}\langle\delta_i\delta_a\rangle-\frac{1}{\HH}b_{L_i}\langle\delta_i\,\partial_r(\bv\cdot\bn)_a\rangle
-\frac{1}{\HH}b_{L_a}\langle\delta_a\,\partial_r(\bv\cdot\bn)_i\rangle+\frac{1}{\HH^2}\langle\partial_r(\bv\cdot\bn)_i\partial_r(\bv\cdot\bn)_a \rangle\Big]\, .\nonumber
\end{align}
Note that if galaxies do not follow Poisson statistics, this contribution will also be modified.

The total variance is then simply the sum of eqs.~\eqref{varP} and~\eqref{varCP}.

\section{Optimising the kernel}
\label{sec:opt}

We want to find the kernel $\w{i}{j}$ which minimises the variance under the constraint that our estimator is unbiased i.e. $\big\langle \hat\xi\big\rangle=\xi_{\rm true}$. This kernel must be symmetric in $i$ and $j$ (see eq.~\eqref{wsym}), since we have used this property to derive the variance. We construct the Lagrangian
\be
\label{L}
L={\rm var}(\hat\xi)+\lambda_0\big[\langle \hat\xi\rangle-\xi_{\rm true} \big]+\sum_{ijL_i L_j}\lambda_{ijL_i L_j}\big(\w{i}{j}- \w{j}{i}\big)\, ,
\ee
where $\lambda_0$ and $\lambda_{ijL_i L_j}$ are Lagrange multipliers. 

Minimising~\eqref{L} with respect to $\lambda_0$ gives $\langle \hat\xi\rangle=\xi_{\rm true}$. Minimising with respect to $\lambda_{cdL^*_c L^*_d}$, where $L^*_c$ and $L^*_d$ denote two specific values of the luminosity in pixels $c$ and $d$, gives $w_{\bx_c \bx_d L^*_c L^*_d}=w_{\bx_d \bx_c L^*_d L^*_c}$. Minimising with respect to 
$w_{\bx_c \bx_d L^*_c L^*_d}$ gives
\be
\label{wcd}
\frac{\partial {\rm var}(\hat\xi)}{\partial w_{\bx_c \bx_d L^*_c L^*_d}}+\lambda_0\frac{\partial \langle \hat\xi\rangle}{\partial w_{\bx_c \bx_d L^*_c L^*_d}}+\lambda_{cdL^*_c L^*_d}-\lambda_{dcL^*_d L^*_c}=0\, ,
\ee
and minimising with respect to $w_{\bx_d \bx_c L^*_d L^*_c}$ gives
\be
\label{wdc}
\frac{\partial {\rm var}(\hat\xi)}{\partial w_{\bx_d \bx_c L^*_d L^*_c}}+\lambda_0\frac{\partial \langle \hat\xi\rangle}{\partial w_{\bx_d \bx_c L^*_d L^*_c}}+\lambda_{dcL^*_d L^*_c}-\lambda_{cdL^*_c L^*_d}=0\, .
\ee
Taking the sum of eqs.~\eqref{wcd} and~\eqref{wdc} we obtain
\be
\label{eqw}
\frac{\partial {\rm var}(\hat\xi)}{\partial w_{\bx_c \bx_d L^*_c L^*_d}}+\frac{\partial {\rm var}(\hat\xi)}{\partial w_{\bx_d \bx_c L^*_d L^*_c}}
+\lambda_0\left( \frac{\partial \langle \hat\xi\rangle}{\partial w_{\bx_c \bx_d L^*_c L^*_d}}+\frac{\partial \langle \hat\xi\rangle}{\partial w_{\bx_d \bx_c L^*_d L^*_c}}\right)=0\, .
\ee 

Inserting the expressions for the mean and the variance into eq.~\eqref{eqw} and dividing by $8d\bar{n}_{L^*_c}d\bar{n}_{L^*_d}$, we can 
rewrite eq.~\eqref{eqw} in matrix notation as
\be
\label{eqwmatrix}
w + wN +N^Tw=B\, ,
\ee
where
\begin{align}
&w_{ij}\equiv \w{i}{j},\hspace{0.9cm} B_{ij}\equiv\frac{\lambda_0}{4}\left(\frac{\dot\HH}{\HH^2}+\frac{2}{r\HH} \right)(b_{L_i}-b_{L_j})\big\langle \delta_i\, (\R)_j\big\rangle\hspace{0.9cm}\mbox{and}\\
&N_{ij}\equiv d\bar n_{L_i}\left[ b_{L_i}b_{L_j}\langle\delta_i\delta_j\rangle-\frac{1}{\HH}b_{L_i}\langle\delta_i\,\partial_r(\bv\cdot\bn)_j\rangle
-\frac{1}{\HH}b_{L_j}\langle\delta_j\,\partial_r(\bv\cdot\bn)_i\rangle+\frac{1}{\HH^2}\langle\partial_r(\bv\cdot\bn)_i\partial_r(\bv\cdot\bn)_j \rangle\right]\, .
\end{align}
To solve eq.~\eqref{eqwmatrix}, we first add the term $N^TwN$. Using the same steps as in appendix~\ref{a:variance} we can indeed show that this term is zero. With this term eq.~\eqref{eqwmatrix} becomes
\be
(\unit+N^T)w(\unit+N)=B\, ,
\ee
and the solution is
\be
w=(\unit+N^T)^{-1}B(\unit+N)^{-1}\, .\label{kernelfinal}
\ee
Eq.~\eqref{kernelfinal} is the kernel which maximises the signal-to-noise for the dipole. 
This kernel is relatively sophisticated since for each pair of pixels it involves a sum over all other pairs of pixels in the survey. Its role is to maximise the signal, while simultaneously minimising the noise due to both Poisson sampling and the product of Poisson sampling with cosmic variance. In~\cite{weightbias}, it has been shown that the measurement of the standard power spectrum can be improved by using bias-dependent weights. In addition~\cite{shotnoise} showed that Poisson noise can be significantly reduced by using an appropriate weighting. Our optimal kernel incorporates these effects in configuration space, for the measurement of the relativistic dipole.

Kernel~\eqref{kernelfinal} can be simplified in the case where the Poisson noise dominates over the mixed term (this is usually the case at small separations, see appendix~\ref{a:B}). In this case, $\unit$ dominates over $N$ and we can expand eq.~\eqref{kernelfinal} in powers of $N$ around $\unit$. We obtain at lowest order
\begin{align}
\label{kernel2}
\w{i}{j}\simeq \frac{\lambda_0}{4}\left(\frac{\dot\HH}{\HH^2}+\frac{2}{r\HH} \right)(b_{L_i}-b_{L_j})\langle \delta_i\, (\R)_j\rangle\,.
\end{align}
This kernel is quite intuitive: it shows that the measurement of the dipole can be improved by weighting more the cross-correlations between galaxies with very different biases, for which the signal is larger. It is however not optimal at large separation, where the mixed term dominates over the Poisson noise. In the following, we will use the simplified kernel~\eqref{kernel2} to explicitly calculate the signal-to-noise. As we will see using this kernel already improves the signal-to-noise significantly. We defer to a forthcoming paper the study of the whole kernel~\eqref{kernelfinal}, which will lead to further improvement, especially at large scales.

\section{Signal-to-noise}
\label{sec:signaltonoise}

We calculate the signal-to-noise for the dipole, using the kernel in eq.~\eqref{kernel1} (kernel 1) and the optimised kernel in eq.~\eqref{kernel2}  (kernel 2). This kernel is defined up to a constant of normalisation which we choose similarly to eq.~\eqref{kernel1}. We can write
\be
\label{kernel_gen}
\w{i}{j}=\frac{3}{8\pi}g(d_{ij})h(L_i, L_j)\cos\beta_{ij} \delta_K(d_{ij}-d)\, ,
\ee
where
\be
\label{defgh}
g(d_{ij})=\left\{\begin{array}{ll} 1 & \quad\mbox{for kernel 1}\\ \alpha(d_{ij})&\quad\mbox{for kernel 2} \end{array}\right.
\hspace{0.8cm} \mbox{and}\hspace{0.8cm} h(L_i, L_j)=\left\{\begin{array}{ll} \theta(L_i-L_j) -\theta(L_j-L_i) & \quad\mbox{for kernel 1}\\ b_{L_i}-b_{L_j}&\quad\mbox{for kernel 2} \end{array}\right.
\ee
with
\be
\alpha(d_{ij})=\frac{1}{2\pi^2}\left(\frac{\dot\HH}{\HH^2}+\frac{2}{r\HH} \right)\frac{\HH}{\HH_0}f
\int dk \,k \HH_0P(k, \bar z)j_1(k d_{ij})\, .
\ee

The mean of the estimator is
\bea
\label{mean_2kernel}
\langle\hat\xi \rangle&=& \frac{3}{8\pi}\sum_{ijL_iL_j}d\bar n_{L_i}d\bar n_{L_j}(b_{L_i}-b_{L_j})g(d_{ij})h(L_i, L_j)\alpha(d_{ij})\cos^2\!\beta_{ij}\delta_K(d_{ij}-d)\, .
\eea
As shown in appendix~\ref{a:continuous}, in the continuous limit this expression reduces to
\bea
\label{mean_cont}
\langle\hat\xi \rangle
&=&\frac{1}{2}N_{\rm tot} \ell_pd^2\bar{N}g(d)\alpha(d)\sum_{L L'}\bar n_{L }\bar n_{L'}h(L, L')(b_{L}-b_{L'})\, ,
\eea
where $\ell_p$ denotes the length of the cubic pixels, $N_{\rm tot}$ is the total number of galaxies in the survey, $\bar N$ is the mean number density:
\be
\bar{N}=\frac{1}{\ell^3_p}\sum_{L}d\bar{n}_{L}\, ,
\ee
and $\bar{n}_{L}$ is the fractional number of galaxies with luminosity $L$:
\be
\bar{n}_{L}=\frac{d\bar{n}_{L}}{\sum_{L'}d\bar{n}_{L'}}\, .
\ee
 In the case of two populations of galaxies we obtain for the kernel 1 and 2
\bea
\langle\hat\xi \rangle_{{\rm K}_1}&=&N_{\rm tot} \ell_pd^2\bar{N}\alpha(d)\bar n_{B}\bar n_{F}(b_{B}-b_{F} )\, ,\label{mean_kernel1}\\
\langle\hat\xi \rangle_{{\rm K}_2}&=&N_{\rm tot} \ell_pd^2\bar{N}\alpha^2(d)\bar n_{B}\bar n_{F}(b_{B}-b_{F} )^2\, .
\eea

The Poisson contribution to the variance is given by (see appendix~\ref{a:continuous} for more detail) 
\bea
{\rm var}_{\rm P}(\hat\xi)&=&2\left(\frac{3}{8\pi} \right)^2 \sum_{ijL_iL_j}d\bar n_{L_i}d\bar n_{L_j}h^2(L_i, L_j)g^2(d_{ij})\cos^2\beta_{ij}\delta_K(d_{ij}-d)\delta_K(d-d')\nonumber\\
&=&\frac{3}{8\pi}N_{\rm tot} \ell_pd^2\bar{N}g^2(d)\sum_{LL'}\bar n_{L}\bar n_{L'}h^2(L, L')\delta_D(d-d')\, . \label{varP_cont}
\eea

The mixed term in the variance is more complicated to calculate since it contains a sum over three pixels. The derivation is presented in appendix~\ref{a:continuous}, where we show that in the continuous limit this term can be written as
\begin{align}
\label{varCP_cont}
{\rm var}_{\rm CP}(\hat\xi)=&\frac{9}{8\pi}N_{\rm tot}\bar{N}^2 \ell_p^2 d^2 d'^2 g(d)g(d')\Big[D_0 \sigma_0(d, d')+D_2\sigma_2(d, d')+D_4\sigma_4(d, d') \Big]\, .
\end{align}
Here
\be
\sigma_\ell(d, d')=-\frac{1}{2\pi^2}\int_{-1}^1 d\mu\, \mu\int_{-1}^1 d\nu\, \nu\int_0^{2\pi}d\varphi\int dk k^2 P(k, \bar z) j_\ell(ks)P_\ell\left(\frac{d\mu+d'\nu}{s}\right)\, ,\hspace{0.3cm}\ell=0,2,4\, ,
\ee
and
\be
s=\sqrt{d^2+d'^2+2dd'\big(\mu\nu+\sqrt{(1-\mu^2)(1-\nu^2)}\sin\varphi \big)}\, .
\ee
The coefficients $D_\ell$ are defined as
\begin{align}
D_0&=\sum_{LL'L''}h(L,L')h(L'',L')\bar n_L \bar n_{L'}\bar n_{L''}\left[b_L b_{L''}+(b_L+b_{L''})\frac{f}{3}+\frac{f^2}{5} \right]\, ,\label{D0}\\
D_2&=-\sum_{LL'L''}h(L,L')h(L'',L')\bar n_L \bar n_{L'}\bar n_{L''}\left[(b_L+b_{L''})\frac{2f}{3}+\frac{4f^2}{7} \right]\, ,\label{D2}\\
D_4&=\sum_{LL'L''}h(L,L')h(L'',L')\bar n_L \bar n_{L'}\bar n_{L''}\frac{8f^2}{35}\, .\label{D4}
\end{align}
Note that contrary to the Poisson contribution, the mixed term does not vanish for $d\neq d'$. It introduces therefore correlations of the dipole at different separations.  

Combining eqs.~\eqref{mean_cont}, \eqref{varP_cont} and \eqref{varCP_cont}, the signal-to-noise at fixed separation $d$ becomes
\begin{align}
\label{SNfin}
\frac{S}{N}(d)=\frac{\langle\hat\xi \rangle(d)}{\sqrt{{\rm var}(\hat\xi)}(d)}=\sqrt{\frac{2\pi N_{\rm tot}}{3}}
\frac{A\,\alpha(d)}
{\left[B\left(\bar N d^2 \ell_p \right)^{-1}
+3 \Big(D_0\sigma_0(d, d)+D_2\sigma_2(d, d)+D_4\sigma_4(d, d)\Big)\right]^{1/2}}\, ,
\end{align}
where
\bea
A&=&\sum_{LL'}\bar n_{L}\bar n_{L'}h(L, L')(b_{L}-b_{L'})\,,\label{A}\\
B&=&\sum_{LL'}\bar n_{L}\bar n_{L'}h^2(L, L')\, .\label{B}
\eea
These coefficients as well as the $D_\ell$ defined in eqs.~\eqref{D0} to~\eqref{D4} depend on the biases and number densities of the different populations of galaxies that we are cross-correlating. In the case where we have only two populations of galaxies, the two kernels give exactly the same signal-to-noise since the function $h(L,L')$ can be factorised out of eqs.~\eqref{D0}-\eqref{D4}, \eqref{A} and \eqref{B}.

If the noise is dominated by Poisson sampling, the signal-to-noise~\eqref{SNfin} becomes
\be
\frac{S}{N}= \frac{A\,\alpha(d)}{\sqrt{6B}}\sqrt{N_{\rm pair}}\, ,
\ee
where $N_{\rm pair}(d)=4\pi N_{\rm tot}\bar N\ell_p d^2$ is the total number of pairs at fixed separation $d$. This number depends on the size of the pixel $\ell_p$.

On the other hand, if the noise is dominated by the mixed term, the signal-to-noise is
\begin{align}
\frac{S}{N}=\sqrt{2\pi N_{\rm tot}}\,
\frac{A\,\alpha(d)}
{3\left[\Big(D_0\sigma_0(d, d)+D_2\sigma_2(d, d)+D_4\sigma_4(d, d)\Big)\right]^{1/2}}\, .
\end{align}
In this case, the signal-to-noise depends only on the total number of galaxies $N_{\rm tot}$ and is independent on the pixelisation.

Eq.~\eqref{SNfin} represents the signal-to-noise at fixed separation $d$. To calculate the cumulative signal-to-noise over a range of separation, we need to account for the covariances between separations. We have
\be
\label{SNtot}
\left(\frac{S}{N} \right)_{\rm cum}^2=\sum_{ij}\langle\hat\xi \rangle(d_i){\rm var}(\hat\xi)^{-1}(d_i, d_j)\langle\hat\xi \rangle(d_j)\, ,
\ee
where ${\rm var}(\hat\xi)$ is given by the sum of eqs.~\eqref{varP_cont} and \eqref{varCP_cont}.

\section{Forecasts}

\label{sec:measurement}

\begin{figure}[t]
\centering
\includegraphics[width=0.465\textwidth]{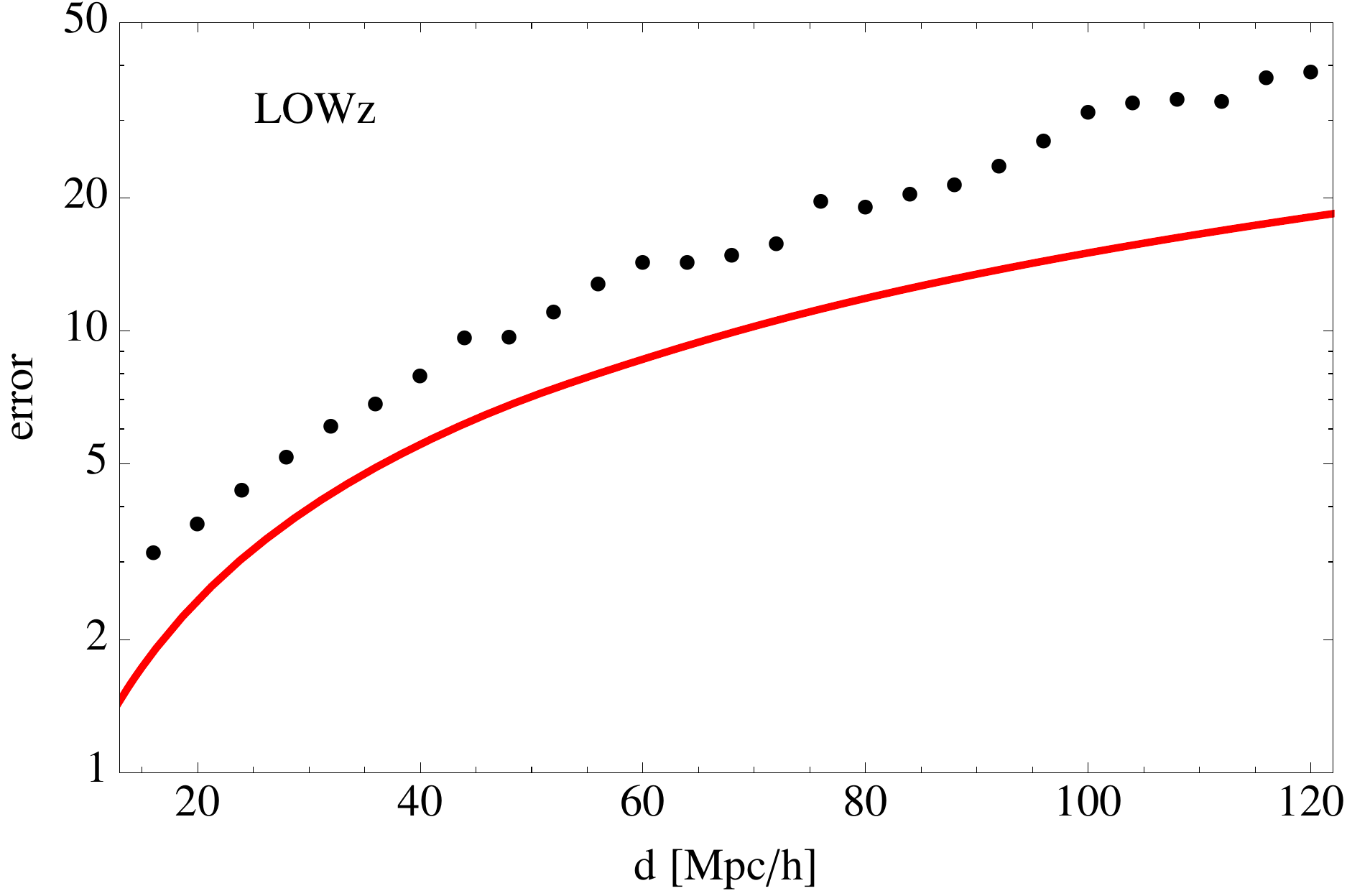}\hspace{0.8cm}\includegraphics[width=0.482\textwidth]{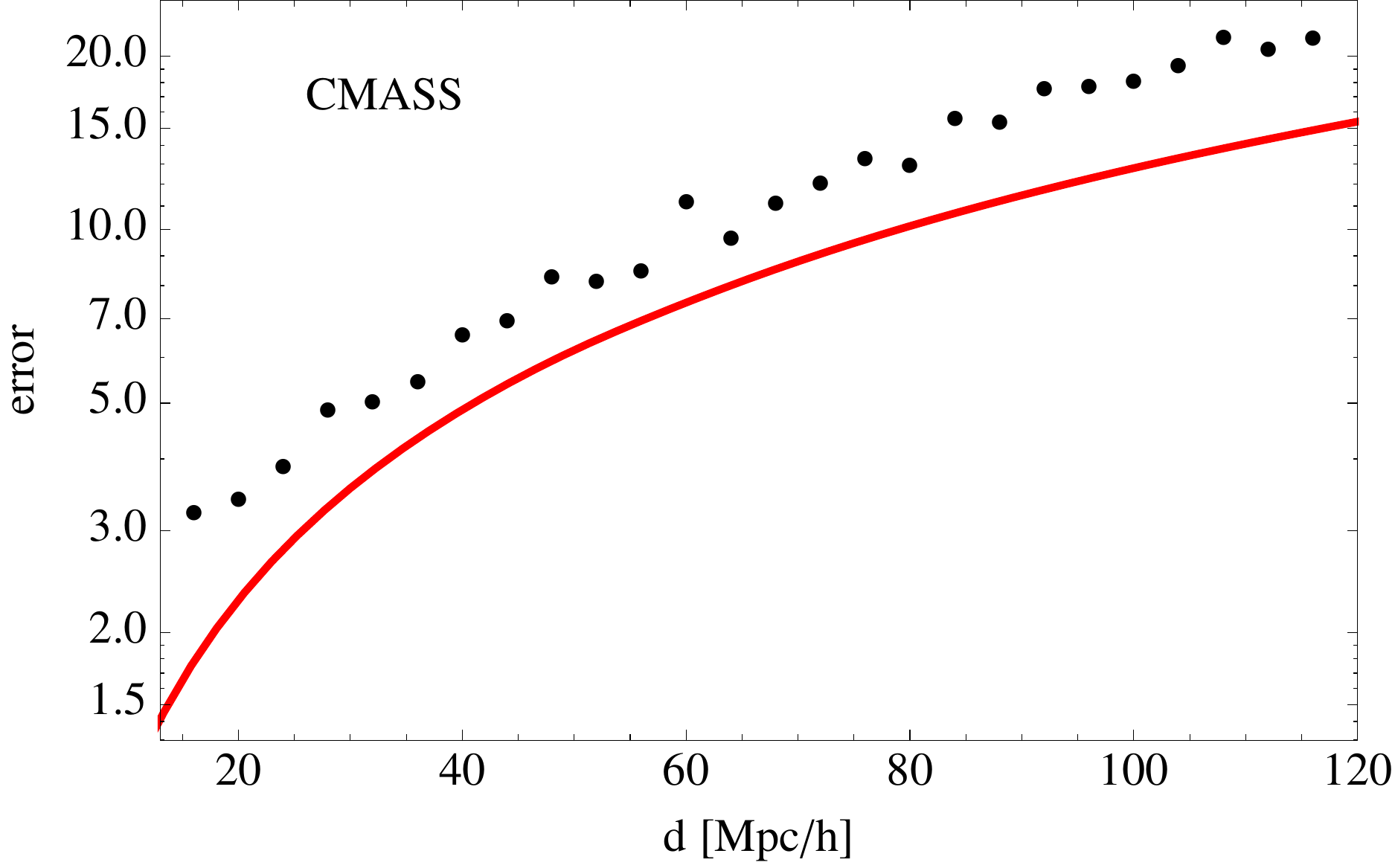}
\caption{\label{fig:error} Errors in the measurements of the dipole (black dots) in the LOWz and CMASS samples. The red solid line represents the theoretical prediction from eqs.~\eqref{varP_cont} and \eqref{varCP_cont}, calculated with pixel size $\ell_p=4$\,Mpc/$h$.}
\end{figure}

We apply now our method to concrete examples. We start by validating the calculation of the errors using measurements from the BOSS survey. In~\cite{kernel_obs} we present a measurement of the dipole for two populations of galaxies in the LOWz and CMASS samples, data release DR10~\cite{bossDR11}, using kernel 1. Here we use this measurement to compare the observed errors (obtained by Jackknife) with our predictions using eqs.~\eqref{varP_cont} and \eqref{varCP_cont}.
Since the number density of galaxies as well as the fractional number of bright and faint galaxies are evolving through the sample, we split them into thin redshift slices of $\Delta z\sim0.01$ and we calculate the errors in each bin. The total error in the samples is then obtained by averaging the errors over the redshift slices~\footnote{Note that to compare the theoretical signal to the measurement, we must divide eq.~\eqref{mean_kernel1} by $N_{\rm tot} \ell_pd^2\bar{N}\bar n_{B}\bar n_{F}$. We therefore apply the same normalisation to eqs.~\eqref{varP_cont} and \eqref{varCP_cont}.}. The mean bias of the bright and faint populations have been measured using the monopole of the total sample and the monopole of the cross-correlation between the bright and faint populations~\cite{kernel_obs}. In LOWz we found a mean bias for the bright population of $b_\B=2.30$ and for the faint population of $b_\F=1.31$. In the CMASS sample we found $b_\B=2.36$ and $b_\F=1.46$. In both samples we use cubic pixels with size $\ell_p=4$\,Mpc/$h$ for separations $16\leq d\leq 120$\,Mpc/$h$. 
The errors are shown in figure~\ref{fig:error}. Our prediction under-estimates the Jackknife errors by up to 30 percent both in the LOWz and in the CMASS sample. We have checked that the Jackknife errors for the monopole and the quadrupole agree well with the errors from the BOSS collaboration~\cite{bossDR11}, obtained from simulations. This gives us confidence that the Jackknife errors on the dipole are reliable. The 30 percent difference between the Jackknife errors and our theoretical predictions could be due to inhomogeneous sampling in the BOSS data. These inhomogeneities are accounted for by weighting the data appropriately and are consequently captured by the Jackknife errors but they are not encoded in the theoretical predictions. We will investigate this in more detail in a future work.

\subsection{Multi-population of halos in the millennium simulation}
\label{sec:millennium}

We now calculate the signal-to-noise for multiple populations of galaxies in three concrete examples. First we use the millennium-XXL simulation~\cite{millenniumXXL}. The millennium simulation does not contain relativistic distortions and it is therefore not expected to contain a dipole~\footnote{Note that some simulations, like for example the MICE simulation~\cite{MICE}, do provide the distribution of galaxies {\it on the light-cone}. As a consequence, those simulations contain part of the relativistic dipole, namely all the terms proportional to the velocity in eq.~\eqref{Deltarel}.}. This simulation is however useful since it contains various populations of halos with different masses (and consequently biases) that we can use as a model to predict how the optimal kernel increases the signal-to-noise.

We use measurements from~\cite{millpop}, where halos in the millennium simulation have been separated into 6 mass bins. For each bin, the bias and number density have been measured. These measurements are summarised in table~\ref{table:millpop}. They were performed at $z=0$. Here we use them to calculate the signal-to-noise for the dipole between $z=0$ and $z=0.2$, assuming that they are valid over this redshift range. We split the volume into two redshift bins of width $\Delta z=0.1$ and we calculate the signal-to-noise in each bin. Since the redshift bins are to a good approximation uncorrelated we obtain the total signal-to-noise by adding in quadrature the signal-to-noise from each bin.
The volume of the box is $(4.11\, {\rm Gpc}/h)^3$. In this box we count the number of halos we would see on our past light-cone in each of the redshift bins. The number density in the simulation is $\bar N=2.7\times 10^{-3}\, (h/{\rm Mpc})^3$ leading to a total number of halos  $N_{\rm tot}=2.9 \times 10^5$ in the lowest redshift bin and $N_{\rm tot}=1.86 \times 10^6$ in the highest redshift bin.

\begin{center}
\begin{table}[h]
\begin{tabular}{| c | c | c | }
\hline
Halo mass \hspace{0.05cm} [$M_\odot h^{-1}$]&\hspace{0.1cm} $b_L$ \hspace{0.1cm}& \hspace{0.1cm}$\bar n_L$ \hspace{0.1cm}\\
\hline \hline
$1-3\times 10^{12}$& 0.8 & 0.842 \\
$7-8\times 10^{12}$& 1& 0.045\\
$1-3\times 10^{13}$ & 1.1&0.092\\
$5-7\times 10^{13}$ & 1.5&0.009\\
$9\times 10^{13}-3\times 10^{14}$ & 2 &0.01\\
$3-6\times 10^{14}$ & 2.9&0.002\\
\hline
\end{tabular}
\caption{\label{table:millpop} Biases and fractional number densities for six populations of halos with different masses, measured in the millennium simulation at $z=0$~\cite{millpop}.}
\end{table}
\end{center}

\begin{figure}[t]
\centering
\includegraphics[width=0.5\textwidth]{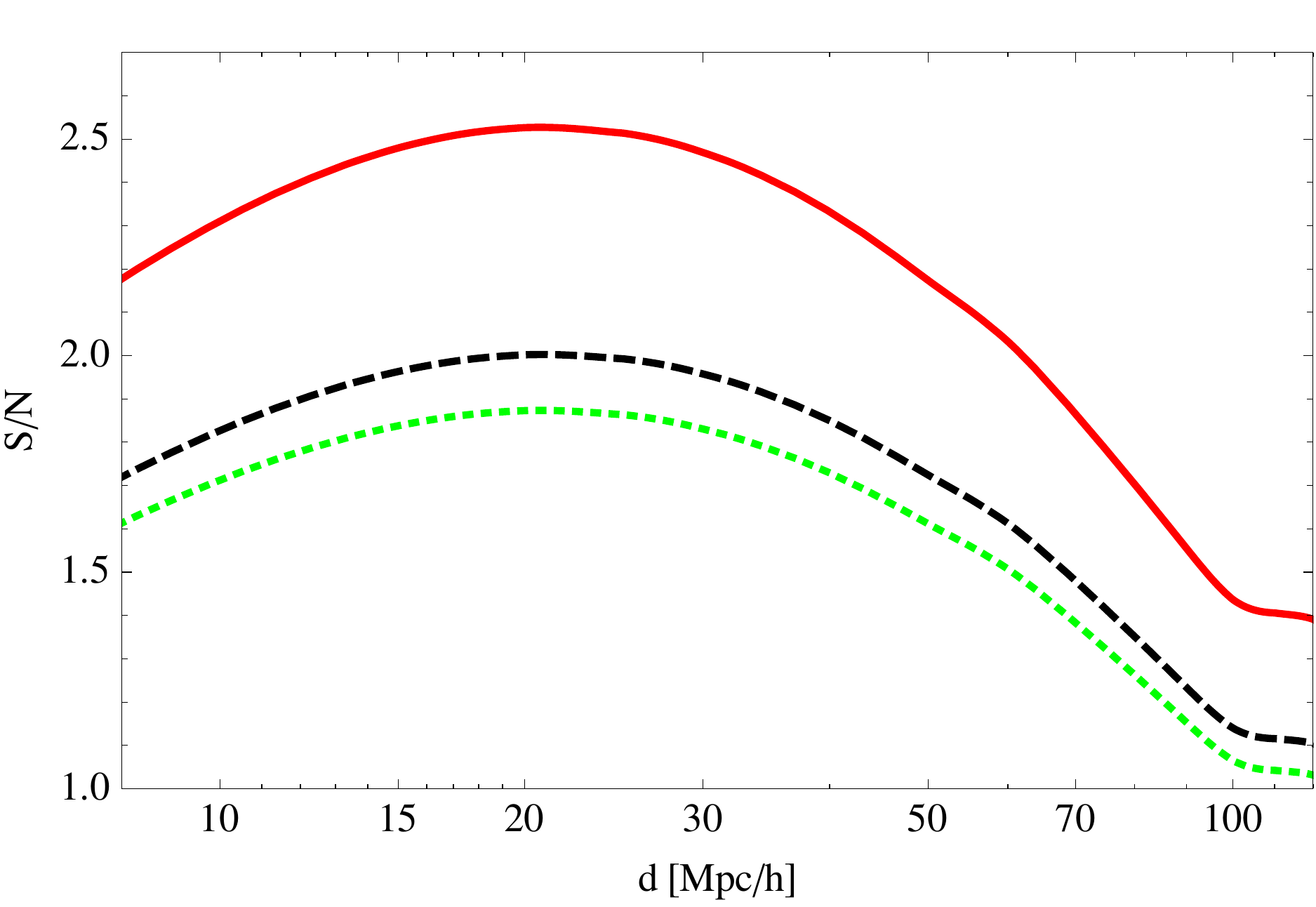}
\caption{\label{fig:SN_mill} Predicted signal-to-noise for the dipole between $z=0$ and $z=0.2$ in a survey with the same characteristics as the millennium simulation, plotted as a function of separation between galaxies. We use pixels of size $8\,{\rm Mpc}/h$. The green dotted line corresponds to two populations with kernel 1 (same with kernel 2). The black dashed line corresponds to six populations with kernel 1 and the red solid line to six populations with kernel 2.}
\end{figure}

We calculate the signal-to-noise in three different cases: the six populations of table~\ref{table:millpop} with kernel 1; the six populations of table~\ref{table:millpop} with kernel 2; and two populations, the first one (faint) corresponding to the first mass bin in table~\ref{table:millpop}, and the second one (bright) combining bins 2 to 6. The mean bias for the bright population is $b_\B=1.17$ and the mean fractional number of galaxies is $\bar n_\B=0.158$. The signal-to-noise for the three cases is plotted in figure~\ref{fig:SN_mill}. Going from two populations of galaxies to six populations of galaxies increases the signal-to-noise by 7 percent. This is simply due to the fact that for six populations the number of possible cross-correlations is larger than for two populations and the noise is consequently smaller. Then using the optimal kernel provides a further improvement of 28 percent, leading to a total improvement of 35 percent. 

The cumulative signal-to-noise from 8 to 120\,Mpc/$h$ can be calculated from eq.~\eqref{SNtot}. For two populations of galaxies we find a cumulative signal-to-noise of 3.3. With six populations and kernel 1 the signal-to-noise becomes 3.5, whereas the optimal kernel allows us to reach a cumulative signal-to-noise of 4.4. This indicates that with the optimal kernel a detection of the dipole in a survey with characteristics similar to those of the millennium simulation should be possible. Note that the naive cumulative signal-to-noise obtained by simply summing over separations without accounting for the covariance between bins is of 7.8 instead of 4.4 for the optimal kernel. This shows that the bins in separation are significantly correlated.

\subsection{Multi-populations of galaxies in the main sample of SDSS DR5}
\label{sec:DR5}

As a second example we apply our method to the main sample of galaxies in the data release DR5 of SDSS. This sample has two advantages with respect to the BOSS LOWz and CMASS samples. First it is at lower redshift, where the dipole is larger. And second it contains a more diverse population of galaxies for which the biases are significantly different. In~\cite{DR5pop2}, Percival et al. split the main galaxy sample into nine bins of luminosity and measured the bias for each population. For our calculation of the signal-to-noise we group these nine populations into six populations. The mean number density of galaxies is $\bar N=4.3\times 10^{-3}\, (h/{\rm Mpc})^3$. As before we split the survey into two redshift bins of width $\Delta z=0.1$ and we calculate the signal-to-noise in each of the bins. In total we have 465'789 galaxies: 62'705 in the lowest redshift bin and 403'083 in the highest redshift bin. The values of the biases and fractional number densities for the different populations are extracted from~\cite{DR5pop2, DR5pop} and are summarised in table~\ref{table:DR5pop}. 

\begin{center}
\begin{table}[h]
\begin{tabular}{| c | c | c | }
\hline
Mean magnitude&\hspace{0.1cm} $b_L$ \hspace{0.1cm}& \hspace{0.1cm}$\bar n_L$ \hspace{0.1cm}\\
\hline \hline
-22.5& 2.16 & 0.046 \\
-21.75& 1.68& 0.017\\
-21.25 & 1.44&0.017\\
-20.25 & 1.32&0.328\\
-19.25 & 1.08 &0.164\\
-18.5 & 0.96&0.428\\
\hline
\end{tabular}
\caption{\label{table:DR5pop} Biases and fractional number densities for six populations of galaxies with different magnitude, measured in the data release DR5 of SDSS~\cite{DR5pop, DR5pop2}.}
\end{table}
\end{center}

\begin{figure}[t]
\centering
\includegraphics[width=0.5\textwidth]{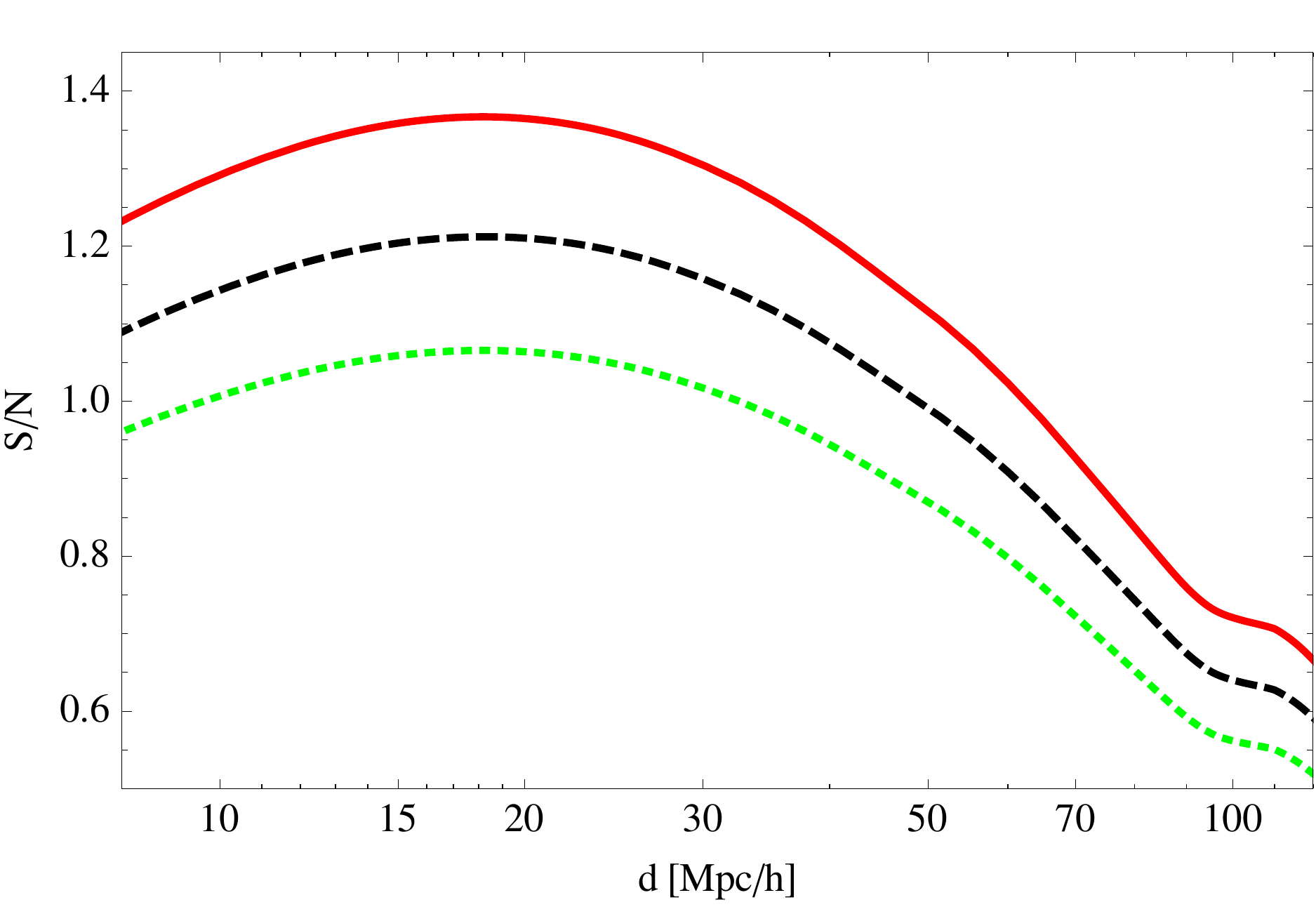}
\caption{\label{fig:SN_DR5} Predicted signal-to-noise for the dipole in the data release DR5 of SDSS $z\leq 0.2$, plotted as a function of separation between galaxies. We use pixels of size $8\,{\rm Mpc}/h$. The green dotted line corresponds to two populations with kernel 1 (same with kernel 2). The black dashed line corresponds to six populations with kernel 1 and the red solid line to six populations with kernel 2.}
\end{figure}

We calculate the signal-to-noise in three different cases: the six populations of table~\ref{table:DR5pop} with kernel 1; the six populations of table~\ref{table:DR5pop} with kernel 2; and two populations, the first one (bright) combining the bins 1 to 4 in table~\ref{table:millpop}, and the second one (faint) combining the bins 5 and 6. The mean bias for the bright population is $b_\B=1.43$ and the mean fractional number of galaxies is $\bar n_\B=0.408$. The mean bias for the faint population is $b_\F=0.99$ and the mean fractional number of galaxies is $\bar n_\F=0.592$. The signal-to-noise for the three cases is plotted in figure~\ref{fig:SN_DR5}. The optimal kernel with six populations increases the signal-to-noise by 28 percent with respect to the case of two populations: we gain 14 percent by going from two populations to six populations with kernel 1, and using kernel 2 gives a further increase of 14 percent. 
The cumulative signal-to-noise from 8 to 120\,Mpc/$h$ is of 1.8 in the case of two populations, 2.1 with six populations and kernel 1 and 2.4 with six populations and the optimal kernel.
 
\subsection{The Dark Energy Spectroscopic Instrument}
\label{sec:DESI}

\begin{figure}[t]
\centering
\includegraphics[width=0.5\textwidth]{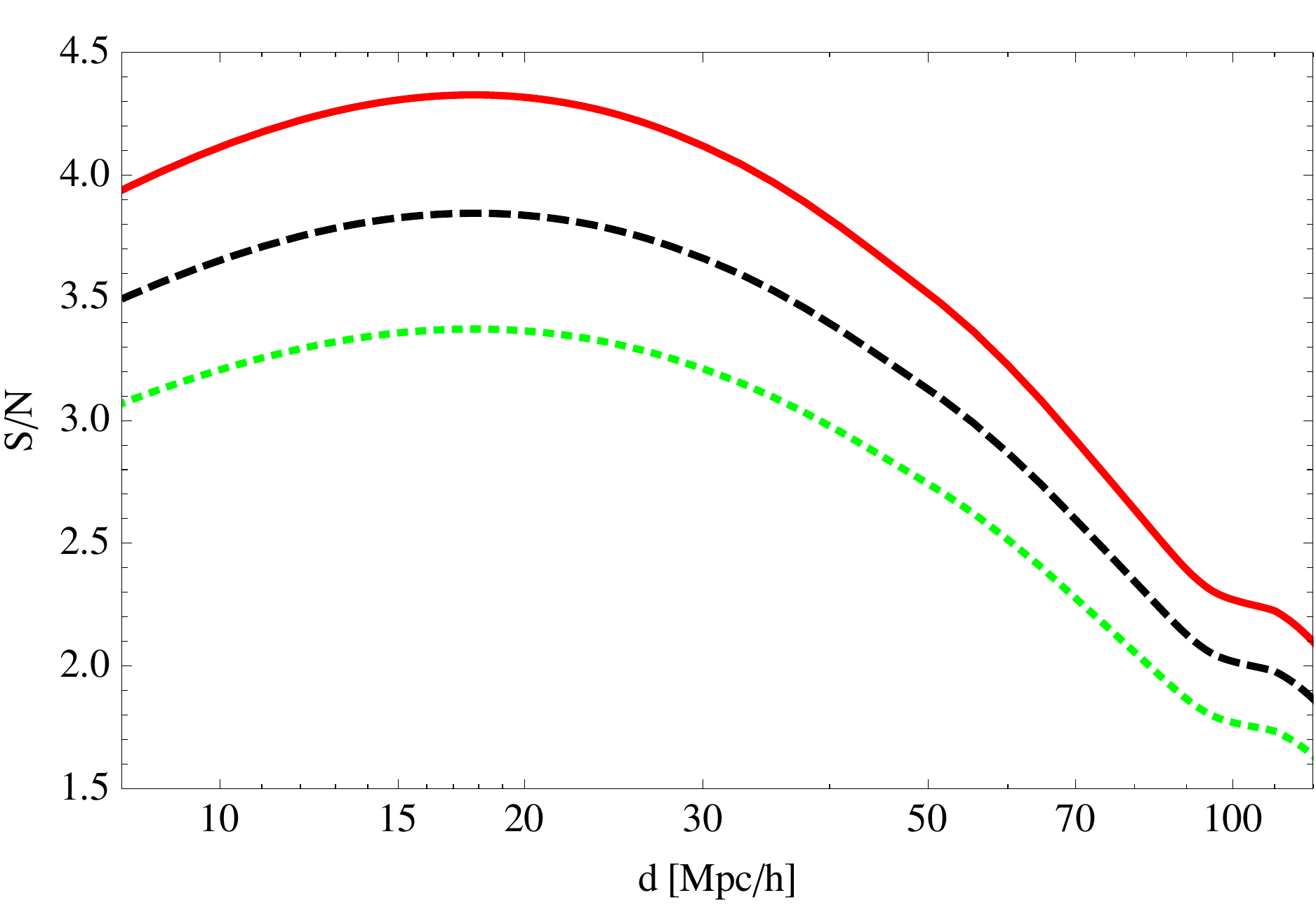}
\caption{\label{fig:SN_DESI} Predicted signal-to-noise for the dipole in the DESI Bright Galaxy sample $z\leq 0.3$, plotted as a function of separation between galaxies. We use pixels of size $8\,{\rm Mpc}/h$. The green dotted line corresponds to two populations with kernel 1 (same with kernel 2). The black dashed line corresponds to six populations with kernel 1 and the red solid line to six populations with kernel 2.}
\end{figure}

Finally we forecast the signal-to-noise for the future Dark Energy Spectroscopic Instrument (DESI)~\cite{desi2}. The Bright Galaxy DESI survey~\cite{desi} will observe 10 million galaxies at low redshift $z\leq0.3$ over 14'000 square degrees. We split this sample into three redshift bins of size $\Delta z=0.1$ and calculate the signal-to-noise in each bin using table~\ref{table:DR5pop}, i.e. assuming that the DESI Bright Galaxy sample can be split in luminosity in a similar way as the SDSS sample. The signal-to-noise in shown in figure~\ref{fig:SN_DESI}. With 6 populations and the optimal kernel we reach a signal-to-noise of 4.3 per bin between 15 and 30\,Mpc/$h$. The improvement of the signal-to-noise with respect to SDSS is mainly due to the fact that DESI will observe a significantly larger number of galaxies, thanks to the fact that both the volume and the galaxy number density are larger in DESI than in SDSS. As shown in eq.~\eqref{SNfin} the signal-to-noise is indeed proportional to $\sqrt{N_{\rm tot}}$. This increases the signal-to-noise by a factor 2.6 in the redshift bins $0<z<0.1$ and $0.1<z<0.2$. In addition DESI will observe 6.8 million galaxies at redshift $0.2<z<0.3$ leading to a total improvement of a factor 3.1 with respect to SDSS. The cumulative signal-noise from 8 to 120\,Mpc/$h$ is of 5.8 in the case of two populations, 6.6 with six populations and kernel 1 and 7.4 with six populations and the optimal kernel. The optimal kernel increases therefore the signal-to-noise by 28 percent with respect to the two populations case. Note that these numbers may be reduced by $\sim$ 30 percent, if the errors in DESI are affected by inhomogeneous weights as is the case for BOSS (see figure~\ref{fig:error}). 
These forecasts seem consistent with the results of~\cite{yoo_im}, who calculated the detectability of relativistic effects using the power spectrum of multiple populations of galaxies. For a full-sky survey at low redshift $0.1<z<0.3$ they predict a 5-sigma detection of the imaginary part of the power spectrum if all halos down to bias of order 1 are used (corresponding to a minimum halo mass of $3\times 10^{11}\,M_\odot h^{-1}$).

In addition to the Bright sample, DESI will observe emission line galaxies (ELG) and bright luminous red galaxies (LRG) over a wide range of redshift. 
We use the specifications of~\cite{desi2} (see table 3) to forecast the signal-to-noise from the two lowest redshift bins, i.e. from $z=0$ to $z=0.4$. We use the ELG's as faint sample with bias $b_\F=0.84$, and the LRG's as bright sample with bias $b_\B=1.7$. The cumulative signal-to-noise is of 4.6.
Since ELG's and LRG's have very different biases, we can expect to split the sample into more populations following table~\ref{table:millpop}. According to figure~\ref{fig:SN_mill} this could increase the signal-to-noise by up to 35 percent giving a cumulative signal-to-noise of 6.2. These forecasts show that a robust detection of the relativistic dipole should be possible in the near future.

\section{Conclusions}
\label{sec:conclusion}

Relativistic distortions are an intrinsic part of our observations. They are rich in information since they are sensitive not only to the galaxies' peculiar velocities, but also to the geometry of the universe through the metric potentials $\Phi$ and $\Psi$. Measuring relativistic distortions would therefore open the way to new tests of the theory of gravity.  These effects are however challenging to detect since they are suppressed by powers of $\HH/k$ with respect to the standard contributions. To observe the impact of relativistic distortions on the monopole, quadrupole and hexadecapole, we need therefore to look at correlation functions at horizon scales $k\sim \HH$.  

In this paper we propose instead to isolate the relativistic distortions by fitting for a dipole in the cross-correlation function between multiple populations of galaxies. 
The advantage of using the dipole to measure relativistic distortions is twofolds. First, this allows us to remove the contribution from the standard terms, which in the distant-observer approximation affect only the even multipoles. Second, the kernel to isolate the dipole is anti-symmetric in the exchange of the galaxies' luminosity, which automatically suppresses the cosmic variance contribution to the error. 

Combining multiple populations of galaxies, we construct an optimal estimator to maximise the signal-to-noise of the dipole. This estimator has a complicated form, which involves multiple summations over all pixels in the survey, as shown in eq.~\eqref{kernelfinal}. In a forthcoming paper we will study how to implement efficiently this estimator in large-scale structure surveys. Here instead we restrict ourself to the case where Poisson sampling dominates the error. In this case, the estimator takes a simple and intuitive form, which can readily be used. We find that with this simple estimator we increase the signal-to-noise of the dipole by up to 35 percent. This allows us to reach a detectable level for the dipole in the main SDSS sample of galaxies and in surveys with number densities and halos similar to the one in the millennium simulation. In a forthcoming paper we will apply this method to the data release DR5 of SDSS and to the MICE simulation. This will require to split the galaxies into multiple populations with different luminosities and to combine these populations according to eq.~\eqref{mean_2kernel}. In the future we will also try and measure the dipole in the upcoming DESI survey for which we predict a cumulative signal-to-noise of 7.4.   

A detection of the relativistic dipole with this method would be very interesting since it would allow us to test the validity of Euler equation in a model independent way. According to eq.~\eqref{Deltarel} the dipole is indeed sensitive to a combination of the gravitational potential $\Psi$ and the peculiar velocity of galaxies. Combining a measurement of the dipole with a measurement of the quadrupole would therefore allow us to test the relation between $\Psi$ and $V$, i.e. to test if the velocity of galaxies is governed only by the gravitational potential as predicted in General Relativity, or if a fifth force acts on the galaxies.

\[\]
{\bf Acknowledgments:} It is a pleasure to thank Alex Hall for interesting discussions and for his help in showing the cancellation of the cosmic variance. We also thank Rupert Croft, Ofer Lahav and Francesco Montanari for stimulating and useful discussions. We thank the referee for his useful comments which help us improve the manuscript. CB acknowledges support by the Swiss National Science Foundation. LH is supported in part by the DOE and NASA.
EG acknowledges support from projects AYA2012-39559 and  Consolider-Ingenio CSD2007- 00060 
from he Spanish Ministerio de Ciencia e Innovacion (MICINN) and CosmoComp (PITNGA-2009-238356) and 
LACEGAL (PIRSES-GA-2010-269264) from the European Commission.

\appendix

\section{Comparison of $N$ and $\unit$ in eq.~\eqref{kernelfinal}}
\label{a:B}

In eq.~\eqref{kernelfinal} we have to compare the two terms $w_{ij}$ and $\sum_a w_{ia} N_{aj}$. For simplicity we fix the position of the pixels $i$ and $j$ on the $z$ axis and we look at a bright galaxy in pixel $i$ and a faint galaxy in pixel $j$. Neglecting the effect of redshift-space distortions, we need to compare $\frac{3}{8\pi}\cos\beta_{ij}=\frac{3}{8\pi}$ with 
\be
I_{ij}\equiv \sum_a w_{ia} N_{aj}=\frac{3}{8\pi}d\bar n_\F b_\F^2\sum_a \cos\beta_{ia}C_0(d_{aj})\, .
\ee
In the continuous limit we obtain
\be
I_{ij}=\frac{3}{4}\bar n_\F b_\F^2 \bar N \int_{-1}^1 d\mu \mu \int_0^{\infty} ds s^2 C_0\big(d_{ij}^2+s^2-2d_{ij}s\mu\big)\, .
\ee
$I_{ij}$ can be calculated numerically for fixed values of $d_{ij}$. Choosing $\bar N=2.8\times 10^{-4}$, which is the number density in the LOWz survey, we find that at small separation $d_{ij}=1\,{\rm Mpc}/h$, $I_{ij}= 0.033$ and at large separation, $d_{ij}=50\,{\rm Mpc}/h$, $I_{ij}= 0.74$. Comparing this with $\frac{3}{8\pi}=0.12$, we see that at small separation $w_{ij}$ dominates over $\sum_a w_{ia} B_{aj}$, whereas at large separation it is the opposite. For surveys with larger number density, like the main galaxy sample of DR5, $\sum_a w_{ia} B_{aj}$ starts dominating over $w_{ij}$ at small separation already. 

\section{Explicit calculation of the mean and variance in the continuous limit}
\label{a:continuous}

\subsection{Mean}

Inserting the kernel~\eqref{kernel_gen} into eq.~\eqref{xirel} we find for the mean of the estimator
\bea
\langle\hat\xi \rangle&=& \frac{3}{8\pi}\sum_{ijL_iL_j}d\bar n_{L_i}d\bar n_{L_j}(b_{L_i}-b_{L_j})g(d_{ij})h(L_i, L_j)\alpha(d_{ij})\cos^2\!\beta_{ij}\delta_K(d_{ij}-d)\nonumber\, .
\eea
The functions $d\bar n_{L_i}$, $b_{L_i}$ and $\alpha$ depend on the redshift $z_i$. However we neglect here the evolution of these functions with redshift, meaning that we evaluate them at the mean redshift of the survey $\bar z$ (or the mean of the redshift slice). Since these functions evolve slowly with redshift and since in the distant-observer approximation we have $|z_i-z_j|\ll \bar z$, this is a good approximation. In this case, the sum over luminosities $L_i$ and $L_j$ is independent of the pixels and can be written as
\be
\sum_{L_iL_j}d\bar n_{L_i}d\bar n_{L_j}h(L_i, L_j)(b_{L_i}-b_{L_j})=\sum_{LL'}d\bar n_{L}d\bar n_{L'}h(L, L')(b_{L}-b_{L'})\, .
\ee
In the continuous limit, the sum over pixels becomes 
\be
\sum_i =\frac{1}{\ell^3_p}\int d^3\bx\hspace{0.7cm}\mbox{and}\hspace{0.7cm}\delta_K(d_{ij}-d)=\ell_p\delta_D(|\bx-\by|-d)\, ,
\ee
where $\ell_p$ denotes the size of the cubic pixels. With this we obtain
\be
\langle\hat\xi \rangle=\frac{3}{8\pi}\frac{\ell_p}{\ell_p^6}\sum_{LL'}d\bar n_{L}d\bar n_{L'}h(L, L')(b_{L}-b_{L'})
\int d^3\bx \int d^3\by\, g(|\bx-\by|)\alpha(|\bx-\by|)\cos^2\!\beta(\bx,\by)\delta_D(|\bx-\by|-d)\, .
\ee
We can fix the position of $\bx$ and integrate over all $\by$. Since the integrand does not depend on the position of $\bx$ but only on the relative separation $|\bx-\by|$ and on the orientation of the pair with respect to the observer $\beta(\bx,\by)$, the integral over $\bx$ simply gives the volume of the survey $V$ (or the volume of the redshift bin over which we average). We obtain
\bea
\langle\hat\xi \rangle&=& \frac{3}{8\pi}\frac{\ell_pV}{\ell_p^6}\sum_{LL'}d\bar n_{L}d\bar n_{L'}h(L, L')(b_{L}-b_{L'})
\int_0^{2\pi}d\varphi \int_0^\pi d\beta \sin\beta \cos^2\beta \int_0^\infty \!ds\, s^2 g(s)\alpha(s) \delta_D(s-d)\nonumber  \\
&=&\frac{1}{2}N_{\rm tot}\ell_pd^2\bar{N}g(d)\alpha(d)\sum_{L L'}\bar n_{L }\bar n_{L'}h(L, L')(b_{L}-b_{L'})\, ,
\eea
where 
\be
\bar{N}=\frac{1}{\ell^3_p}\sum_{L}d\bar{n}_{L}
\ee
is the mean number density in the survey,
\be
\bar{n}_{L}=\frac{d\bar{n}_{L}}{\sum_{L'}d\bar{n}_{L'}}
\ee
is the fractional number of galaxies with luminosity $L$, and $N_{\rm tot}=\bar N V$ is the total number of galaxies in the survey. 

\subsection{Variance}
\label{a:variance}

\noindent{\bf Poisson term}

\hspace{0.2cm}

The Poisson contribution to the variance is
\be
{\rm var}_{\rm P}(\hat\xi)=2\left(\frac{3}{8\pi} \right)^2 \sum_{ijL_iL_j}d\bar n_{L_i}d\bar n_{L_j}h^2(L_i, L_j)g^2(d_{ij})\cos^2\beta_{ij}\delta_K(d_{ij}-d)\delta_K(d-d')\, .
\ee
In the continuous limit we obtain
\bea
{\rm var}_{\rm P}(\hat\xi)&=&2\left(\frac{3}{8\pi} \right)^2\frac{\ell_p}{\ell_p^6}
\sum_{L L'}d\bar n_{L}d\bar n_{L'}h^2(L, L')
\int d^3\bx \int d^3\by\, g^2(|\bx-\by|)\cos^2\!\beta(\bx,\by)\delta_D(|\bx-\by|-d)\delta_D(d-d')\nonumber\\
&=&\frac{3}{8\pi}N_{\rm tot}\ell_pd^2\bar{N}g^2(d)\sum_{LL'}\bar n_{L}\bar n_{L'}h^2(L, L')\delta_D(d-d')\, .
\eea

\vspace{5cm}

\noindent{\bf Cosmic variance}

\vspace{0.2cm}

The cosmic variance contribution is given by
\begin{align}
\label{var_cosmic}
{\rm var_C}(\hat\xi)=&2\sum_{ijab}\sum_{L_iL_jL_a L_b}d\bar n_{L_i}d\bar n_{L_j}d\bar n_{L_a}d\bar n_{L_b}\w{i}{j}\w{a}{b}\\
&\times\Big[ b_{L_i}b_{L_a}\langle\delta_i\delta_a\rangle-\frac{1}{\HH}b_{L_i}\langle\delta_i\,\partial_r(\bv\cdot\bn)_a\rangle
-\frac{1}{\HH}b_{L_a}\langle\delta_a\,\partial_r(\bv\cdot\bn)_i\rangle+\frac{1}{\HH^2}\langle\partial_r(\bv\cdot\bn)_i\partial_r(\bv\cdot\bn)_a \rangle\Big]\nonumber\\
&\times\Big[ b_{L_j}b_{L_b}\langle\delta_j\delta_b\rangle-\frac{1}{\HH}b_{L_j}\langle\delta_j\,\partial_r(\bv\cdot\bn)_b\rangle
-\frac{1}{\HH}b_{L_b}\langle\delta_b\,\partial_r(\bv\cdot\bn)_j\rangle+\frac{1}{\HH^2}\langle\partial_r(\bv\cdot\bn)_j\partial_r(\bv\cdot\bn)_b \rangle\Big]\, .\nonumber
\end{align}
We can easily show that many of the products in the brackets vanish since they are symmetric under the exchange $L_i\leftrightarrow L_j$ or $L_a\leftrightarrow L_b$, whereas the kernels $\w{i}{j}$ and $\w{a}{b}$ are anti-symmetric. The remaining terms read

\begin{align}
{\rm var_C}(\hat\xi)=&\sum_{ijab}\sum_{L_iL_jL_a L_b}d\bar n_{L_i}d\bar n_{L_j}d\bar n_{L_a}d\bar n_{L_b}\w{i}{j}\w{a}{b}
(b_{L_i}-b_{L_j})(b_{L_a}-b_{L_b})\\
&\times  \frac{1}{\HH^2}\Big[\langle\delta_i\delta_a \rangle \langle\partial_r(\mathbf{V}\cdot\mathbf{n})_j\partial_r(\mathbf{V}\cdot\mathbf{n})_b \rangle
- \langle\delta_i\partial_r(\mathbf{V}\cdot\mathbf{n})_a \rangle\langle\delta_j\partial_r(\mathbf{V}\cdot\mathbf{n})_b \rangle \Big]\, .\nonumber
\end{align}
Using that
\begin{align}
\langle\delta_i\delta_j \rangle&=\frac{1}{(2\pi)^3}\int d^3\bk \,e^{i \bk (\bx_j-\bx_i)}P(k)\, ,\\
-\frac{1}{\HH}\langle\delta_i\partial_r(\mathbf{V}\cdot\mathbf{n})_j \rangle&=\frac{f}{(2\pi)^3}\int d^3\bk \,e^{i \bk (\bx_j-\bx_i)}P(k)
\left[\frac{1}{3}+\frac{2}{3}P_2(\bn \cdot \hat{\bk})\right]\, ,\\
\frac{1}{\HH^2}\langle\partial_r(\mathbf{V}\cdot\mathbf{n})_i\partial_r(\mathbf{V}\cdot\mathbf{n})_j \rangle&=
\frac{f^2}{(2\pi)^3}\int d^3\bk \,e^{i \bk (\bx_j-\bx_i)}P(k)\left[\frac{1}{5}+\frac{4}{7}P_2(\bn \cdot \hat{\bk})
+\frac{8}{35}P_4(\bn \cdot \hat{\bk})\right]\, ,
\end{align}
where $P_\ell$ denotes the Legendre polynomial of degree $\ell$,
and going to the continuous limit we find
\begin{align}
\label{varPl}
{\rm var_C}(\hat\xi)=&\frac{4 f^2}{(2\pi)^6}\frac{1}{\ell_p^{12}}\sum_{LL'L'' L'''}d\bar n_{L}d\bar n_{L'}d\bar n_{L''}d\bar n_{L'''}(b_{L}-b_{L'})(b_{L''}-b_{L'''})\\
&\times \int d^3 \bx_i \int d^3 \bx_j \int d^3 \bx_a \int d^3 \bx_b\, w_{\bx_i\bx_jL L'}w_{\bx_a\bx_bL'' L'''} 
\int d^3\bk \int d^3\bk' \,e^{i\bk(\bx_a-\bx_i)}e^{i\bk'(\bx_b-\bx_j)}\nonumber\\
&\times P(k)P(k')\left[\frac{1}{45}+\frac{2}{63}P_2(\bn \cdot \hat{\bk}')+\frac{2}{35}P_4(\bn \cdot \hat{\bk}')
-\frac{1}{9}P_2(\bn \cdot \hat{\bk}')P_2(\bn \cdot \hat{\bk}) \right]\, .\nonumber
\end{align}

We then do the following change of variables $\bx_i\rightarrow \by_i=\bx_j-\bx_i$ and $\bx_a\rightarrow\by_a=\bx_b-\bx_a$.
The exponentials become
\be
e^{i\bk(\bx_a-\bx_i)}e^{i\bk'(\bx_b-\bx_j)}=e^{i\bk(\by_i-\by_a)}e^{i(\bk+\bk')(\bx_b-\bx_j)}\, .
\ee
The kernel $\w{i}{j}$ is a function of the separation between the pixels $|\bx_j-\bx_i|=y_i$ and the orientation of the pair with respect to the line-of-sight $\cos\beta_{ij}=\cos\beta_{y_i}$, and similarly for $\w{a}{b}$.
The integral over $\bx_j$ and $\bx_b$ become then trivial
\begin{align}
&\int d^3 \bx_j \, e^{-i(\bk+\bk')\bx_j}=(2\pi)^3 \delta_D(\bk+\bk') \quad \mbox{and} \quad \int d^3 \bx_b =V\, ,
\end{align}
where $V$ is the volume of the survey. The Dirac-delta function enforces $\bk'=-\bk$ which implies that the square bracket in eq.~\eqref{varPl} exactly vanishes
\begin{align}
\frac{1}{45}+\frac{2}{63}P_2(\bn\cdot\hat{\bk})+\frac{2}{35}P_4(\bn\cdot\hat{\bk})-\frac{1}{9}P_2^2(\bn\cdot\hat{\bk}) =0\, .
\end{align}
This shows that the measurement of the relativistic dipole is not affected at all by the cosmic variance of the density and redshift-space distortions. This cancellation of the cosmic variance for multiple populations of galaxies has already been demonstrated for the case of the power spectrum~\cite{seljak_tracers,mcdonald_tracers}.

\hspace{0.5cm}

\noindent{\bf Mixed term}

\hspace{0.2cm}

\begin{figure}[!t]
\centering
\includegraphics[width=0.7\textwidth]{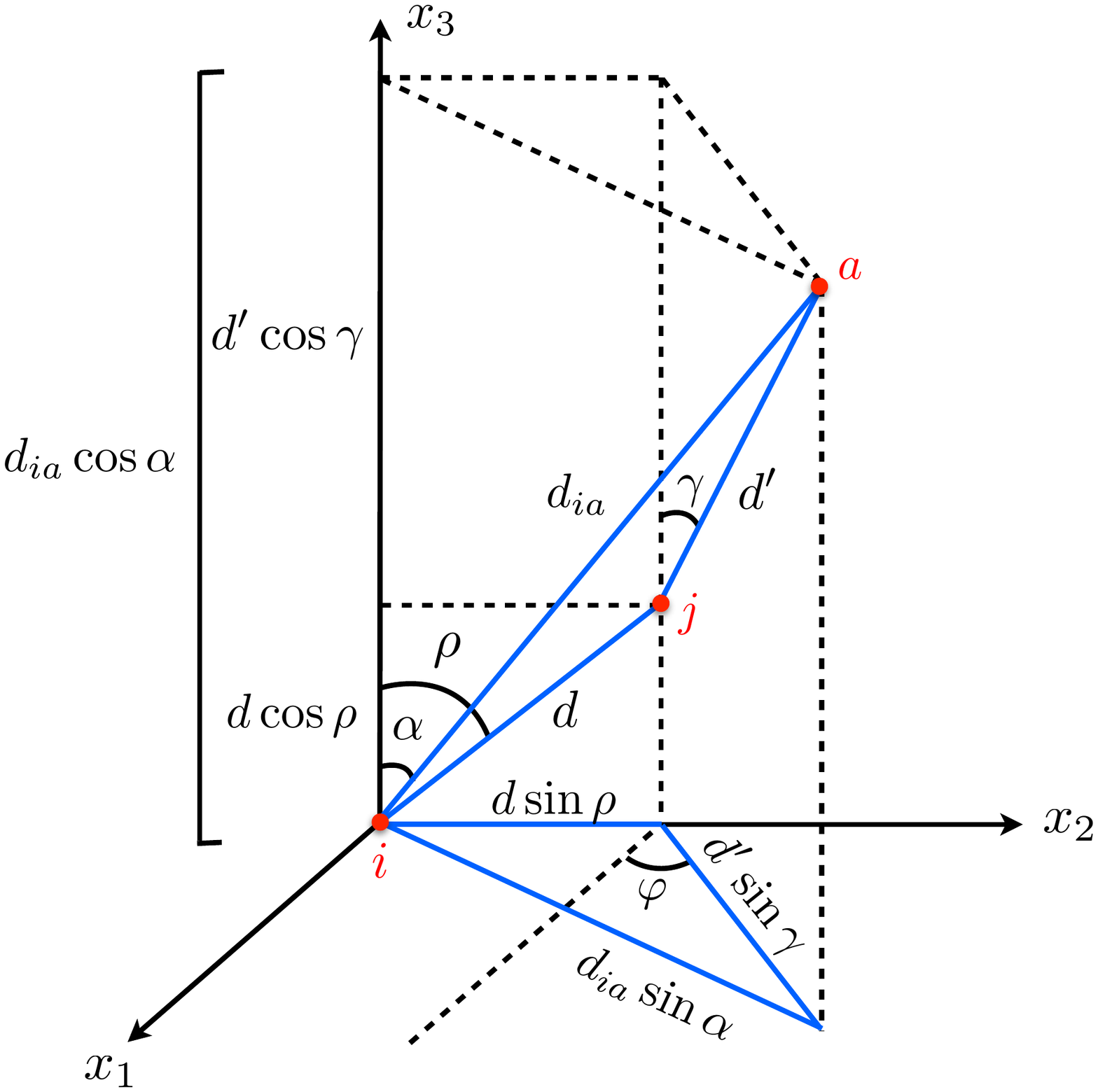}
\caption{\label{fig:H} Configuration used to calculate the mixed term in the variance, eq.~\eqref{varCP_app}. The direction of observation $\bn$ is along the $x_3$ axis.}
\end{figure}

The variance due to the product of the Poisson contribution and of the cosmic variance contribution is
\begin{align}
{\rm var}_{\rm CP}(\hat\xi)=&4\left(\frac{3}{8\pi}\right)^2\sum_{ija}
g(d_{ij})g(d_{aj})\cos\beta_{ij}\cos\beta_{aj}\delta_K(d_{ij}-d)\delta_K(d_{aj}-d')\label{varCP_app}\\
&\times\sum_{L_i L_j L_a}d\bar n_{L_i}d\bar n_{L_j}d\bar n_{L_a}h(L_i, L_j)h(L_a, L_j)\Bigg\{\left[b_{L_i}b_{L_a}+(b_{L_i}+b_{L_a})\frac{f}{3}+\frac{f^2}{5} \right] C_0(d_{ia}) \nonumber\\ 
& -\left[(b_{L_i}+b_{L_a})\frac{2f}{3}+\frac{4f^2}{7} \right] C_2(d_{ia})P_2(\cos\beta_{ia})+\frac{8f^2}{35}C_4(d_{ia})P_4(\cos\beta_{ia}) \Bigg\}\, , \nonumber
\end{align}
where
\bea
C_\ell(d_{ia})&=&\frac{1}{2\pi^2}\int dk k^2 P(k, \bar z) j_\ell(kd_{ia})\, ,\quad \ell= 0, 2, 4\, .
\eea 
Eq.~\eqref{varCP_app} contains a sum over three pixels, which becomes a triple integral in the continuous limit. 
To solve this triple integral, we fix $i$ at the origin and we fix $j$ in the plane $x_2-x_3$ as shown on figure~\ref{fig:H}. The direction of observation $\bn$ is along the $x_3$ axis. Due to the symmetry of the problem we can then simply multiply the result 
by the volume of the survey $V$ (to account for the integral over $i$) and by $2\pi$ to account for the integral over $j$ around the $x_3$ axis. We obtain
\begin{align}
{\rm var}_{\rm CP}(\hat\xi)=&-\frac{9}{8\pi}N_{\rm tot}\bar{N}^2 \ell_p^2 d^2d'^2 g(d)g(d')\int_{-1}^1 d\mu\, \mu\int_{-1}^1 d\nu\, \nu\int_0^{2\pi}d\varphi\\
&\times\sum_{LL'L''}h(L,L')h(L'',L')\bar n_L \bar n_{L'}\bar n_{L''}\Bigg\{\left[b_L b_{L''}+(b_L+b_{L''})\frac{f}{3}+\frac{f^2}{5} \right]C(d_{ia})\nonumber\\
&- \left[(b_L+b_{L''})\frac{2f}{3}+\frac{4f^2}{7} \right]C_2(d_{ia})P_2(\cos\beta_{ia})+\frac{8f^2}{35}C_4(d_{ia})P_4(\cos\beta_{ia})\Bigg\}\, ,\nonumber
\end{align}
where $\mu=\cos\rho$ and $\nu=\cos\gamma$. 
The distance $d_{ia}$ and the angle $\beta_{ia}$ are functions of $d, d', \mu, \nu$ and $\varphi$. From figure~\ref{fig:H} we have
\bea
d_{ia}\cos\alpha&=&d'\cos\gamma+d\cos\rho\\
d_{ia}^2\sin^2\alpha&=&d'^2\sin^2\gamma+d^2\sin^2\rho-2dd'\sin\gamma\sin\rho\cos(\varphi+\pi/2)\, ,
\eea
leading to
\bea
d_{ia}^2&=&d^2+d'^2+2dd'\big(\mu\nu+\sqrt{(1-\mu^2)(1-\nu^2)}\sin\varphi \big)\, ,\\
\cos\beta_{ia}&=&\cos\alpha=\frac{d\mu+d'\nu}{d_{ia}}\, .
\eea


\end{document}